\documentclass{IEEEtran}
\usepackage{cite}
\usepackage{amsmath,amssymb,amsfonts}
\usepackage{algorithmicx}
\usepackage{algorithm}
\usepackage[noend]{algpseudocode}
\makeatletter
\def\BState{\State\hskip-\ALG@thistlm}
\usepackage{graphicx}
\usepackage{textcomp}
\usepackage{xcolor}
\usepackage{setspace}
\def\BibTeX{{\rm B\kern-.05em{\sc i\kern-.025em b}\kern-.08em
		T\kern-.1667em\lower.7ex\hbox{E}\kern-.125emX}}
\usepackage{subfigure}
\newtheorem{my_theorem}{Theorem}

\newtheorem{my_lemma}{Lemma}
\newtheorem{my_corollary}{Corollary}
\newtheorem{my_proposition}{Proposition}

	\title{Random Access Protocols for Cell-Free Wireless Network Exploiting Statistical Behavior of THz Signal  Propagation }
\author{Pranay Bhardwaj,~\IEEEmembership{Graduate Student Member,~IEEE}, S.~M.~ Zafaruddin,~\IEEEmembership{Senior Member,~IEEE}, and Amir Leshem,~\IEEEmembership{Fellow,~IEEE}
		\thanks{This work was supported	in part by the Science and Engineering Research Board (SERB), Department of Science and Technology (DST), Government of India, through the Mathematical Research Impact Centric Support (MATRICS) scheme under Grant MTR/2021/000890.}
	\thanks{ A part of this paper with the results of the generalized statistical model with random atmospheric absorption has been submitted for presentation in 2024 IEEE Wireless Communications and Networking Conference (IEEE WCNC 2024) \cite{Bhardwaj2024_wcnc}.}

	\thanks{Pranay Bhardwaj (p20200026@pilani.bits-pilani.ac.in) and S.~M.~Zafaruddin (syed.zafaruddin@pilani.bits-pilani.ac.in)  are  with  the Department of Electrical and Electronics Engineering, Birla Institute of Technology and Science, Pilani, Pilani Campus-333031, Rajasthan, India.}
	
	\thanks{Amir Leshem is with Faculty of Engineering, Bar-Ilan University, Ramat Gan 52900, Israel (email: leshema@biu.ac.il ).}

}

\begin{document}
	\maketitle


%
\begin{abstract}
  The current body of research on terahertz (THz) wireless communications predominantly focuses on its application for single-user backhaul/fronthaul connectivity at sub-THz frequencies. First, we develop a  generalized statistical model for signal propagation at THz frequencies encompassing physical layer impairments, including random path-loss with  Gamma distribution for the molecular absorption coefficient, short-term fading characterized by the  $\alpha$-$\eta$-$\kappa$-$\mu$ distribution, antenna misalignment errors, and transceiver hardware impairments.  Next, we propose random access protocols for a cell-free wireless network, ensuring successful transmission for multiple users with limited delay and energy loss, exploiting the combined effect of random atmospheric absorption, non-linearity of fading, hardware impairments, and antenna misalignment errors. We consider two schemes: a fixed transmission probability (FTP) scheme where the transmission probability (TP) of each user is updated at the beginning of the data transmission and an adaptive transmission probability (ATP) scheme where the TP is updated with each successful reception of the data. We analyze the performance of both protocols using delay, energy consumption, and outage probability with scaling laws for the transmission of a data frame consisting of a single packet from users at a predefined quality of service (QoS).

\end{abstract}

\section{Introduction}
\label{sec:intro}
Terahertz (THz) communication is gaining rapid momentum as a promising technology for the upcoming generation of wireless networks. With its generous unlicensed bandwidth spanning from $0.1$ to $10$ THz, the THz band enables the delivery of high data rates, while simultaneously ensuring low latency and enhanced security for backhaul/fronthaul communications \cite{Koenig_2013_nature,Dang_2020_nature}. THz wireless technology has the potential to play a crucial role in access networks, particularly in dense network deployments with limited coverage range like cell-free networks 
\cite{Ngo2017, Interdonato2019,Zhang2019, Elhoushy2022}.
In particular, cell-free networks are envisioned as a key network architecture in wireless communication, where multiple access points (APs) are strategically deployed within a specific area, serving multiple users in close proximity.  As a consequence, the distance between users and the APs is significantly reduced compared to traditional cellular systems. This inherent characteristic makes the THz band an ideal candidate for integration into multiuser cell-free networks 
\cite{Faisal2020_thz_cell_free, Sayyari2021_thz_cell_free, Abbasi2022_thz_cell_free, Mukherjee2022_thz_cell_free}.

THz wireless backhaul/fronthaul networks have been investigated specifically in the context of single-user transmission \cite{Li2022_dual_hop_thz_fso,Singya2022_hybrid_fso_thz_backhaul,Pai2021_dual_hop_THz_backhaul,Li_2021_THz_AF,Bhardwaj2022_multihop}, limited research have focused on the multi-user THz wireless communications \cite{Ning2021_thz_multiuser, Chen2023_thz_multiuser, Wang2023_thz_multiuser, Chen2020_thz_multiuser, Lee2022_thz_random_access}. The authors in \cite{Ning2021_thz_multiuser}, the authors propose hybrid beamforming applicable to situations involving single-users as well as multiple users. The authors in \cite{Chen2023_thz_multiuser} introduced a communication system for a multi-user scenario in the terahertz range utilizing a uniform circular array incorporating both downlink and uplink transmission strategies. In \cite{Wang2023_thz_multiuser}, the authors propose a sensing-aided THz wideband hybrid precoding for multi-user environment. The authors in \cite{Chen2020_thz_multiuser} address the channel estimation problem for an reconfigurable intelligent surface (RIS)-aided THz multi-user multi-input single-output system utilizing a two-stage channel estimation scheme. In \cite{Lee2022_thz_random_access} presents a physical random access channel preamble design for the 6G cellular communication systems in the sub-THz bands.

The use of THz band for cell-free wireless communication is studied in \cite{Faisal2020_thz_cell_free, Sayyari2021_thz_cell_free, Abbasi2022_thz_cell_free, Mukherjee2022_thz_cell_free}. The authors in \cite{Faisal2020_thz_cell_free} and \cite{Sayyari2021_thz_cell_free} list qualitative research directions for the use of THz in the cell-free massive MIMO (mMIMO) systems. The work presented in \cite{Abbasi2022_thz_cell_free} proposes employing the THz spectrum for wireless backhaul connections connecting unmanned aerial vehicle (UAV) base stations (BSs) to central-processing unit (CPU). The authors in \cite{Mukherjee2022_thz_cell_free} explores the vulnerability of THz wireless technology to physical-layer jamming in the framework of cell-free mMIMO systems. 

Multiuser interference cancellation has been studied extensively for cell-free wireless network \cite{Emil2020,Liu2020, Emil2020twc,Zhang2021,Ibrahim2022}. The interference cancellation requires complex signal processing at the CPU and significant overheads. In this context, random access protocol can be a potential alternative \cite{Rivero2005, Baccelli2013, Wang2016, Vukobratovic2020,Zhang2020, Ebrahimi2023}. The authors in \cite{Rivero2005} introduce random access control strategies designed to mitigate collisions and maintain system stability under high traffic loads by limiting the number of transmissions and retransmissions. The authors in \cite{Baccelli2013} present an adaptive protocol design for ad-hoc networks in which mobiles learn the local topology and incorporate this information to adapt their medium access probability (MAP) selection to their local environment. In \cite{Wang2016}, the authors use an adaptive-opportunistic ALOHA media access control protocol for unmanned aerial vehicle-wireless sensor network (UAV-WSN) systems. In \cite{Vukobratovic2020}, the authors propose a random access scheme for an indoor optical wireless communication (OWC) massive Internet of Things (IoT) scenario such that collisions occurring at access points are resolved by employing a centralized interference cancellation that harnesses both spatial and temporal diversity. In \cite{Zhang2020}, the work presents an adaptive transmission protocol to improve the performance of slotted ALOHA, which transmits periodically in different time slots to minimize packet collisions. The work in \cite{Ebrahimi2023} proposes an adaptive random access protocol for massive IoT networks that exploits nonorthogonal multiple access (NOMA) with short-packet transmissions and automatic request and repeat (ARQ) strategy with a limited number of retransmissions.

The effectiveness of a random access protocol is contingent upon the number of users attempting to access the channel. In contrast to RF, THz signal propagation involves greater complexity, including losses attributed to molecular absorption, short-term fading, antenna misalignment errors, and transceiver hardware impairments. To the authors' best knowledge, there is currently no existing research on a random access protocol designed for a cell-free wireless network that leverages the unique characteristics of THz signal propagation.

In this paper, we develop random access protocols for efficient communication in a multi-user cell-free network over the THz band. The main contributions of this study are as follows:
\begin{itemize}
	\item We develop a  generalized statistical model for THz signal propagation, incorporating random path-loss with a Gamma distribution for the molecular absorption coefficient, short-term fading characterized by the $\alpha$-$\eta$-$\kappa$-$\mu$ distribution, as well as accounting for antenna misalignment errors and transceiver hardware impairments.  Note that existing literature assumed a deterministic path-loss model, which is suitable for stable channel environments like backhaul/fronthaul scenarios. However, in the access link, where obstruction (such as human movement and others) between the transmitter and receiver is prevalent, a deterministic path-loss model may not be appropriate.

	\item We propose random access protocols designed for a cell-free wireless network, ensuring successful transmissions for multiple users with minimal delay and energy loss by leveraging the collective impact of random atmospheric absorption, the non-linearity of fading, hardware impairments, and antenna misalignment errors. We assume that each user is assigned a single packet to transmit during data transmission requiring  novel approach for the analysis compared to the conventional analysis of the ALOHA protocol \cite{rom90book}.
	
	\item We consider two schemes: a fixed transmission probability (FTP) scheme where the transmission probability (TP) of each user is updated at the beginning of the data transmission and an adaptive transmission probability (ATP) scheme where the TP is updated with each successful reception of the data.    The FTP implementation is characterized by simplicity, while the ATP scheme demonstrates superior performance, scaling linearly with the number of active users.
\item We analyze the performance of both protocols using delay, energy consumption, and outage probability with scaling laws for the transmission of a data frame consisting of a single packet from users at a predefined quality of service (QoS).  Computer simulations demonstrate the efficacy of the proposed random access schemes and accuracy in performance assessment with the statistical effect of THz propagation for a cell-free network. 
 
\end{itemize}

{\emph{Notations}: The upper incomplete Gamma function is represented as $\Gamma(a,t) = \int_{t}^{\infty} s^{a-1}e^{-s}ds$. We use $G_{p,q}^{m,n}(.|.)$ to denote Meijer's G-function and $H_{p,q}^{m,n}(.|.)$ for Fox's H-function. We denote the Euler-Mascheroni constant, Napier's constant, the exponential function, and the exponential integral as $\gamma$, $\mathrm{e}$, $\exp(\cdot)$, and $E_1(z) = \int_{z}^{\infty} \frac{\exp(-z)}{z}~dz$, respectively. Until and unless specified, logarithmic operations have the natural base $\mathrm{e}$. }

{\emph{Organization}: The paper is organized as follows. Section \ref{sec:sys_model}, describes the system architecture and channel characteristics. In section \ref{sec:chanel_model}, we derive the statistical analysis for the generalized channel model for THz transmission. Section \ref{sec:protocol} describes the proposed random access protocols.  In section \ref{sec:simulation}, we validate the derived analytical results of the proposed protocols and generalized channel model with the help of Monte Carlo simulations under various channel conditions. Finally, the conclusions are presented in section \ref{sec:conc}.}

\section{System Model} \label{sec:sys_model}
\begin{figure}[tp]	
	\centering \hspace{-1.5cm}
	\includegraphics[scale=0.42]{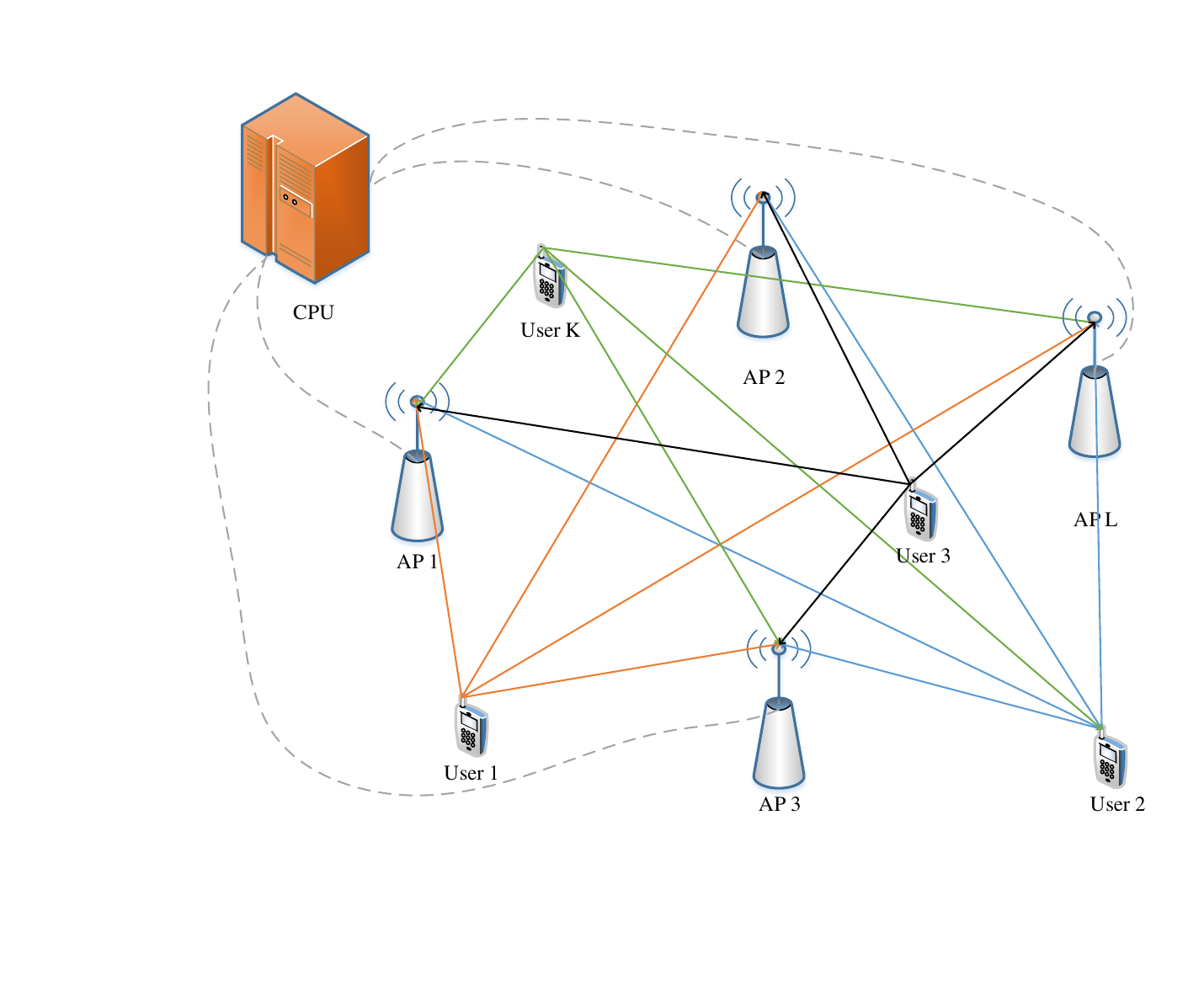}
	\vspace{-1.5cm}
	\caption{System model for a typical cell-free network.}
	\label{fig:sys_mod_twc}	
\end{figure}

As depicted in Fig.~\ref{fig:sys_mod_twc}, we consider a multiuser network consisting of a $ L $ APs that supports $K$ users, all operating on same time-frequency resources within the THz band, emulating a typical cell-free architecture. Consequently, there are $K-1$ interference signals that contribute to the degradation of signal quality for the $i$-th user. 
In this situation, random access methods can provide potential benefits. When utilizing a random access protocol, if the transmission is successful, the signal received at the AP is expressed as \cite{Schenk_hw}:
\begin{flalign} \label{eq:rec_sig}
	y = h(s+w_t) + w_r +w
\end{flalign}
where $h$ represents the channel coefficient, comprising of the path gain $h_l$ due to signal propagation and atmospheric absorption, short-term fading $h_f$, and antenna misalignment errors $h_p$. Here, $s$ denotes the desired signal, and $w$ stands for the additive white Gaussian noise (AWGN) with variance $\sigma_\omega^2$. 

The terms $w_t$ and $w_r$ refer to components representing residual hardware impairment, which are statistically modeled using a Gaussian distribution \cite{Schenk_hw,Boulogeorgos_Analytical}. Specifically, $w_{t}\sim\mathcal{CN}(0,k_{t}^{2}P)$, and $w_{r}\sim\mathcal{CN}(0,k_{r}^{2}P\lvert h \rvert^{2})$, where $k_{t}$ and $k_{r}$ quantify the extent of hardware imperfections in the transmitter and receiver, respectively. In the THz band, typical values of $k_t$ and $k_r$ fall within the range of $(0$-$0.4)$, as suggested in \cite{Boulogeorgos_Analytical}. A value of $k_t=k_r=0$ corresponds to an ideal front-end, representing no hardware imperfections.  The achievable performance is limited by transceiver hardware impairment at higher frequencies  \cite{Pranay_2021_TVT, Pai2021_dual_hop_THz_backhaul, Li_2021_THz_AF, Bhardwaj2022_multihop, Bhardwaj2022_systems_journal, Joshi2022_ftr, Bhardwaj2023_outdoor_thz, Bhardwaj2023_iot,Bhardwaj2023_THI, Boulogeorgos_Analytical}.

The path gain, denoted as $h_l$, relies on various factors such as antenna gains, frequency, and molecular absorption coefficient, as defined in \cite{Boulogeorgos_Analytical}:
\begin{flalign}\label{eq_hl}
	h_l = \frac{c\sqrt{G_{t}G_{r}}}{4\pi f d} \exp(-\frac{1}{2}\kappa(f,T,\psi,p)d)
\end{flalign}
Here, $c$ represents the speed of light, $f$ denotes the transmission frequency, and $d$ represents the distance. $G_{t}$ and $G_{r}$ are the antenna gains of the transmitting and receiving antennas, respectively. The term $\kappa(f,T,\psi,p)$ corresponds to the molecular absorption coefficient, which is influenced by temperature $T$, relative humidity $\psi$, and atmospheric pressure $p$. It can be expressed as:	
\begin{flalign}
	&\kappa(f,T,\psi,p) = \frac{q_1v(q_2v+q_3)}{(q_4v+q_5)^2 + (\frac{f}{100c}-p_1)^2} \nonumber \\ & +  \frac{q_6v(q_7v+q_8)}{(q_9v+q_{10})^2 + (\frac{f}{100c}-p_2)^2} + c_1f^3 + c_2f^2 + c_3f + c_4
\end{flalign}
In the above equation, $v = \frac{\psi}{100} \frac{p_w(T,p)}{p}$, where $p_w(T,p)$ represents the saturated water vapor partial pressure at temperature $T$. The value of $p_w(T,p)$ can be evaluated using Buck's equation.

As is for many wireless communications, short-term fading is inevitable at THz frequencies. Several studies have suggested various distributions to model the short-term fading in the THz band, like $\alpha$-$\mu$, fluctuating two-ray (FTR), and Gaussian mixture distributions  at sub-THz frequencies \cite{Papasotiriou2021_scientific_report}\cite{Papasotiriou2023_scintific_report_outdoor}. The $\alpha$-$\eta$-$\kappa$-$\mu$ model is a comprehensive representation that embodies a wide range of fading characteristics, including the number of multi-path clusters, power of dominant components, nonlinearity of the propagation medium, and scattering level. This flexibility allows the model to effectively capture and fit measurement data in various propagation scenarios, making it an ideal choice for a more generalized and diverse channel model at high frequencies. To model the short-term fading in the THz band for outdoor environment, we adopt the $\alpha$-$\eta$-$\kappa$-$\mu$   fading model \cite{Yacoub_2016_alpha_eta_kappa_mu}.  The authors in \cite{Marins2019_alpha_eta_kapp_mu,Bhardwaj2023_alpha_eta_kappa_mu_globecom} demonstrated that the $\alpha$-$\eta$-$\kappa$-$\mu$  is the best fading model fitting over a wide range of propagation environments  for higher frequency mmWave and THz systems.
We use the exact statistical representation of the $\alpha$-$\eta$-$\kappa$-$\mu$  fading model with PDF of the channel envelope given by \cite{Bhardwaj2023_alpha_eta_kappa_mu_globecom}:
\begin{flalign} \label{eq:pdf_aekm}
&f_{h_f}(h_f) = \frac{\psi_1 \pi^2 2^{(2-\mu)} A_4^{A_1} A_5^{A_2} h_f^{\alpha\mu-1} e^{-\psi_3 h_f^\alpha}}{(\hat{r}^\alpha)^{1+\frac{\mu}{2}}} \nonumber \\ &\times H^{0,1;1,0;1,1;1,0}_{1,1;0,1;2,3;1,3} \Bigg[ \begin{matrix}~ V_1~ \\ ~V_2~ \end{matrix} \Bigg| A_{3}h_f^{\alpha}, \frac{A_4^2}{4}h_f^{\alpha},\frac{A_5^2}{4}h_f^{\alpha}\Bigg],
\end{flalign}
where  $V_1 = \big\{(-A_2;1,0,1)\big\}: \big\{(-,-)\big\} ; \big\{(-A_1,1)(\frac{1}{2},1)\big\} ;\big\{(\frac{1}{2},1)\big\} $ and $V_2 = \big\{(-1-A_1 -A_2;1,1,1) \big\} : \big\{(0,1) \big\} ; \big\{(0,1),(-A_1,1),(\frac{1}{2},1) \big\} ; \big\{(0,1),(-A_2,1),(\frac{1}{2},1) \big\}$, and the constants are defined as $ \psi_1 = \frac{p\alpha\mu^{2}\xi^{1+\frac{\mu}{2}}\delta^{\frac{\mu}{2}-1}q^{\frac{1+p-p\mu}{2+2p}}\eta^{-\frac{1+p+p\mu}{2+2p}}}{\kappa^{\frac{\mu}{2}-1}\exp \left (\frac{(1+pq)\kappa\mu}{\delta}\right)}$, $ \psi_2$ = $\alpha-1$, $ \psi_3= \frac{p\xi\mu}{\eta\hat{r}^\alpha} $, $ A_{1} = \frac{p\mu}{1+p}$-$1, A_{2}= \frac{\mu}{1+p}$-$1, A_{3}= \frac{(\eta-p)\xi\mu}{\eta\hat{r}^\alpha} $, $ A_{4} = 2p\mu \sqrt{\frac{q\kappa\xi}{\eta\delta\hat{r}^\alpha}} $, and $A_{5} = 2\mu\sqrt{\frac{\kappa\xi}{\delta\hat{r}^\alpha}} $. 
Here, $\alpha$ represents the nonlinearity characteristic of the medium, while $\eta$ denotes the ratio of the total power of in-phase and quadrature scattered waves within the multipath clusters. Furthermore, $\kappa$ is defined as the ratio between the total power of dominant components and the total power of scattered waves, and $\mu$ stands for the number of multipath clusters present.

The effect of antenna misalignment errors is also detrimental to THz performance, which occurs when transmit and receive antennas fail to adequately align for line-of-sight (LOS) transmissions, significantly limiting the physical communication range \cite{Dabiri2022_zaf}. The antenna misalignment errors in the THz band for aerial/mobile communication \cite{Dabiri2022_zaf, Badarneh2023_THz_Pointing} are statistically modeled using the PDF
\begin{flalign} \label{eq:pdf_pointing_thz}
	f_{h_{p}}^{}(h_p) = -\rho^2 \ln(h_p) h_p^{\rho-1}
\end{flalign}
where $0<x<1$. $\rho = \sqrt{\frac{w_B^2}{\sigma_\theta^2}}$ determines the severity of the misalignment errors, where $ w_B $ is the angular beamwidth  (beam divergence) of the signal and $ \sigma_\theta $ is the variance of the angular fluctuation, which models the effect of angular fluctuations at both the transmitter and receiver. 

We define the instantaneous SNR of the THz link at the receiver as $\gamma = |h|^2 \bar{\gamma}$, where $h = h_lh_fh_p$ and $\bar{\gamma}$ is the average SNR of the link and is defined as $\bar{\gamma} = \frac{P_t}{\sigma_\omega^2}$ with $P_t$ as the transmit power. From \eqref{eq:rec_sig}, the resultant SNR with transceiver hardware impairment parameters $k_h=\sqrt{k_t^2+k_r^2}$ is given by
\begin{flalign}\label{snr_eq}
	\gamma = \frac{\bar{\gamma} | h_lh_fh_p|^2}{k_h^2 \bar{\gamma} | h_lh_fh_p|^2 +1}
\end{flalign}
It can be seen that SNR $\gamma$ for THz transmission exhibit significant randomness due to the combined effect of $h_l$, $h_f$, and $h_p$. 
In previous studies,  $\kappa(f,T,\psi,p)$ (which determines $h_l$) has been considered deterministic. This assumption holds for backhaul/fronthaul applications where the environment remains relatively stable. However, in cluttered urban environments where factors like human blockage come into play, $\kappa(f,T,\psi,p)$ can exhibit randomness. Further, modeling for $h_f$ becomes more intricate to characterize short-term fading of THz signals. Moreover, there is a randomness in the signal quality due to antenna misalignment parameter $h_p$. 

It is crucial to recognize that when implemented in large multiuser networks, random access protocols can experience drawbacks in terms of latency and inefficient utilization of communication resources. However, unique channel characteristics of THz propagation can enable efficient implementation of the random access protocol.

In what follows, we develop a  statistical model for signal propagation in THz frequencies. This model encompasses random path loss, short-term fading, antenna misalignment errors, and transceiver hardware impairments. Subsequently, we introduce random access protocols designed for a cell-free wireless network. These protocols aim to facilitate successful transmissions for multiple users with minimal delay and energy loss, leveraging the combined impact of random atmospheric absorption, the non-linearity of fading, hardware impairments, and antenna misalignment errors.

\section{A Generalized Channel Model for THz Transmission} \label{sec:chanel_model}
In this section, we develop a generalized channel model for the THz transmission  which includes  several random components for its statistical analysis resulting from atmospheric absorption, misalignment errors, short-term fading, and hardware impairments.

 The THz band experiences higher path-loss due to signal absorption by molecules at extremely small wavelengths \cite{Kim2015, Kokkoniemi_2018, Wu2020}.  The path gain as depicted in \eqref{eq_hl} depends on   molecular absorption coefficient $\zeta(f,T,\psi,p)$.  At sub-THz frequencies, $\zeta(f,T,\psi,p)$ has been considered deterministic. However, at higher frequencies, the interaction of THz signals with the atmosphere can become intricate at the molecular level, making it essential to employ complex modeling for the molecular absorption coefficient $\zeta(f, T, \psi, p)$. As measurement data beyond a few hundred \mbox{GHz} is unavailable for parameterizing $\zeta(f, T, \psi, p)$ at THz frequencies, we resort to a statistical approach. We adopt the Gamma distribution $f_{\zeta(f, T, \psi, p)}(x) = \frac{x^{k-1}}{\beta^k\Gamma(k)} e^{-\frac{x}{\beta}}$ with parameters $k$ and $\beta$ in dB to model $\zeta(f, T, \psi, p)$. This statistical model allows for a diverse range of support for $\zeta(f, T, \psi, p)$ between $0$ and $\infty$, with an average value of $k\beta$ dB/km selected from existing measurement data. In a recent study, the Gamma distribution has been utilized to model the attenuation coefficient in free-space optics (FSO) transmission under foggy weather conditions \cite{fog}. Indeed, experimental data is crucial to validate the use of the Gamma distribution for the absorption coefficient observed in practical scenarios, which presents an excellent opportunity for further research. 
\begin{my_proposition}
	If the molecular absorption coefficient  $\zeta(f, T, \psi, p)$ is Gamma distributed with  parameters $k$ and $\beta$ in dB, then the PDF of the path-gain $h_l$ is given by
	\begin{eqnarray} \label{eq:pdf_path_loss}
	f_{h_{l}}(h_{l}) = \frac{z^k a_l^{-z}}{\Gamma(k)}\bigg[ln\bigg(\frac{a_l}{h_{l}}\bigg)\bigg]^{k-1} h_{l}^{z-1}
	\end{eqnarray}
	where  $a_l = \frac{c\sqrt{G_tG_r}}{4\pi f d}$, $0<h_{l}\le a_l$, and $z = 8.686/(\beta d)$.
\end{my_proposition}

\begin{IEEEproof}
	Converting $\zeta$ in dB, \eqref{eq_hl} can be represented by
	\begin{equation}
	h_l = \frac{c\sqrt{G_{t}G_{r}}}{4\pi f d} \exp\bigg(-\frac{1}{2}\zeta^{\rm dB}(f,T,\psi,p)d^{\rm km}/4.343\bigg)
	\end{equation}
	where $ d^{\rm km} $ is the distance in \mbox{km}. Using $f_{\zeta(f, T, \psi, p)}(x) = \frac{x^{k-1}}{\beta^k\Gamma(k)} e^{-\frac{x}{\beta}}$ and applying standard transformation of random variables, the PDF of $h_l$ is given \eqref{eq:pdf_path_loss}.
\end{IEEEproof}

Next, we present the PDF and CDF of a single link THz transmission, which includes the combined effect of random path loss, antenna misalignment errors, the generalized $\alpha$-$\eta$-$\kappa$-$\mu$ short-term fading, and transceiver hardware impairments. We define $\gamma_h = \sqrt{\frac{\gamma}{\bar{\gamma}(1-\gamma k_h^2)}}$ for the subsequent analysis.
\begin{my_theorem}
	The PDF and CDF of SNR combining the effects of random path loss, antenna misalignment errors, and short-term fading with transceiver hardware impairments for a THz link is given by
	\begin{flalign} \label{eq:pdf_aekm_rpl}
	&f_{\gamma}(\gamma) = \frac{ (-1)^{-k}\psi_1 \rho^2  z^k  }{(\hat{r}^\alpha)^{1+\frac{\mu}{2}} } \bigg(\frac{1}{a_l}\bigg)^{\alpha(1+A_1+A_2)+1}  \nonumber \\ &  \frac{1}{2\big(1-\gamma k_h^2\big)\sqrt{\frac{\bar{\gamma} \gamma}{1-\gamma k_h^2}}} \bigg(\frac{\gamma_h}{\bar{\gamma}}\bigg)^{\frac{\alpha(1+A_1+A_2)}{2}} \nonumber \\ &  H^{0,2+k;1,0;1,1;1,0;1,0}_{2+k,2+k;0,1;2,3;1,3;0,1} \Bigg[ \begin{matrix}~ V_5~ \\ ~V_6~ \end{matrix} \Bigg| \frac{A_{3}\gamma_h^{\frac{\alpha}{2}}}{a_l^\alpha \bar{\gamma}^{\frac{\alpha}{2}}}, \frac{A_4^2 \gamma_h^{\frac{\alpha}{2}}}{4 a_l^\alpha \bar{\gamma}^{\frac{\alpha}{2}}},\frac{A_5^2 \gamma_h^{\frac{\alpha}{2}}}{4 a_l^\alpha \bar{\gamma}^{\frac{\alpha}{2}}},\frac{\psi_3\gamma_h^{\frac{\alpha}{2}}} {a_l^\alpha \bar{\gamma}^{\frac{\alpha}{2}}} \Bigg],
	\end{flalign}
	where $V_5 = \big\{(-A_2;1,0,1,0)\big\}, \big\{\rho-\alpha-\alpha A_1 -\alpha A_2; \alpha, \alpha, \alpha, \alpha\big\}, \big\{z-\alpha-\alpha A_1 -\alpha A_2; \alpha, \alpha, \alpha, \alpha\big\}_k : \big\{(-,-)\big\} ; \big\{(-A_1,1)(\frac{1}{2},1)\big\} ;\big\{(\frac{1}{2},1)\big\}; \big\{-, -\big\} $ and $V_6 = \big\{(-1-A_1 -A_2;1,1,1,0) \big\}, \big\{\rho-1-\alpha-\alpha A_1 -\alpha A_2; \alpha, \alpha, \alpha, \alpha\big\}, \big\{z-1-\alpha-\alpha A_1 -\alpha A_2; \alpha, \alpha, \alpha, \alpha\big\}_k : \big\{(0,1) \big\} ; \big\{(0,1),(-A_1,1),(\frac{1}{2},1) \big\} ; \big\{(0,1),(-A_2,1),(\frac{1}{2},1) \big\} ; \\ \big\{ (0,1)\big\}$.
	
	\begin{flalign} \label{eq:cdf_aekm_rpl}
	&F_{h_{fpl}}(x) = \frac{ \psi_1 \rho^2  z^k  }{(\hat{r}^\alpha)^{1+\frac{\mu}{2}}} \Big(\frac{1}{a_l}\Big)^{\alpha(1+A_1+A_2)+1}  \bigg(\frac{\gamma}{\bar{\gamma}}\bigg)^{\frac{\alpha(1+A_1+A_2) +1}{2}}   \nonumber \\ &  H^{0,3+k;1,0;1,1;1,0;1,0}_{3+k,3+k;0,1;2,3;1,3;0,1} \Bigg[ \begin{matrix}~ V_7~ \\ ~V_8~ \end{matrix} \Bigg| \frac{A_{3}\gamma_h^{\frac{\alpha}{2}}}{a_l^\alpha \bar{\gamma}^{\frac{\alpha}{2}}}, \frac{A_4^2 \gamma_h^{\frac{\alpha}{2}}}{4 a_l^\alpha \bar{\gamma}^{\frac{\alpha}{2}}},\frac{A_5^2 \gamma_h^{\frac{\alpha}{2}}}{4 a_l^\alpha \bar{\gamma}^{\frac{\alpha}{2}}},\frac{\psi_3\gamma_h^{\frac{\alpha}{2}}} {a_l^\alpha \bar{\gamma}^{\frac{\alpha}{2}}} \Bigg],
	\end{flalign}
	
	where $V_7 = \big\{(-A_2;1,0,1,0)\big\}, \big\{\rho-\alpha-\alpha A_1 -\alpha A_2; \alpha, \alpha, \alpha, \alpha\big\}, \big\{z-\alpha-\alpha A_1 -\alpha A_2; \alpha, \alpha, \alpha, \alpha\big\}_k, \big\{-\alpha-\alpha A_1 -\alpha A_2; \alpha, \alpha, \alpha, \alpha\big\} : \big\{(-,-)\big\} ; \big\{(-A_1,1)(\frac{1}{2},1)\big\} ;\big\{(\frac{1}{2},1)\big\}; \big\{-, -\big\} $ and $V_8 = \big\{(-1-A_1 -A_2;1,1,1,0) \big\}, \big\{\rho-1-\alpha-\alpha A_1 -\alpha A_2; \alpha, \alpha, \alpha, \alpha\big\}, \big\{z-1-\alpha-\alpha A_1 -\alpha A_2; \alpha, \alpha, \alpha, \alpha\big\}_k, \big\{-1-\alpha-\alpha A_1 -\alpha A_2; \alpha, \alpha, \alpha, \alpha\big\} : \big\{(0,1) \big\} ; \big\{(0,1),(-A_1,1),(\frac{1}{2},1) \big\} ; \big\{(0,1),(-A_2,1),(\frac{1}{2},1) \big\} ; \\ \big\{ (0,1)\big\}$.
\end{my_theorem}

\begin{IEEEproof}
	The proof is presented in Appendix A.	
\end{IEEEproof}

It can be seen that Theorem 1 results into a $4$-variate Fox's H-function. Note that multivariate Fox's H- function is extensively used to analyze wireless systems for complicated fading channels \cite{Bhardwaj2022_multihop, Du2020_RIS_THz_HW_Impaiment}. However, to simplify further, as a specific instance of Theorem 1, we consider the special case where $\eta=1$ and $\kappa=0$, which  leads to the  $\alpha$-$\mu$ model \cite{Papasotiriou2021_scientific_report}. In this context, we provide  statistical results in terms of single-variate Fox's H-function:
\begin{my_corollary} \label{cor:pdf_alpha_mu}
	The PDF and CDF of the THz link with random path loss, antenna misalignment errors, and transceiver hardware impairments with $\alpha$-$\mu$ short-term fading are given by
	\small
	\begin{flalign} \label{eq:pdf_hfpl}
	&f_{\gamma}(\gamma) = \frac{\alpha \mu^{\mu} \rho^2 z^k }{ \Omega^{\alpha\mu}\Gamma (\mu)  a_l^{\alpha\mu-z}} \frac{1}{2\big(1-\gamma k_h^2\big)\sqrt{\frac{\bar{\gamma} \gamma}{1-\gamma k_h^2}}} \bigg(\sqrt{\frac{\gamma_h}{\bar{\gamma}}}\bigg)^{\alpha\mu-1} \nonumber \\ & H^{3+k,0}_{2+k,3+k}\Bigg[ \frac{\mu \gamma_h^{\frac{\alpha}{2}}}{\Omega^{\alpha\mu} a_l^\alpha \bar{\gamma}^{\frac{\alpha}{2}}} \Bigg| \begin{matrix}  (1-\alpha\mu+\rho,\alpha)_2, (-\alpha\mu+z+1,\alpha)_k  \\ (0,1), (-\alpha\mu+\rho,\alpha)_2, (-\alpha\mu+z,\alpha)_k	\end{matrix}\Bigg] 
	\end{flalign}
	\begin{flalign} \label{eq:cdf_hfpl}
	&F_{\gamma}(\gamma) = \frac{\alpha \mu^{\mu} \rho^2 z^k }{ \Omega^{\alpha\mu}\Gamma (\mu)  a_l^{\alpha\mu-z}} \bigg(\sqrt{\frac{\gamma_h}{\bar{\gamma}}}\bigg)^{\alpha\mu}  H^{3+k,1}_{3+k,4+k} \nonumber \\ & \Bigg[ \frac{\mu \gamma_h^{\frac{\alpha}{2}}}{\Omega^{\alpha\mu} a_l^\alpha \bar{\gamma}^{\frac{\alpha}{2}}} \Bigg| \begin{matrix} (1-\alpha\mu,\alpha),  (1-\alpha\mu+\rho,\alpha)_2, (-\alpha\mu+z+1,\alpha)_k  \\ (0,1), (-\alpha\mu+\rho,\alpha)_2,  (-\alpha\mu+z,\alpha)_k, (-\alpha\mu,\alpha)	\end{matrix}\Bigg] 
	\end{flalign}
\end{my_corollary}
\normalsize
\begin{IEEEproof}
	The proof follows similar steps as that of Theorem 1.
\end{IEEEproof}

In the following corollary, we  simplify the statistical outcomes by assuming that short-term fading can be disregarded. This scenario might arise in specific situations, such as when dealing with a shorter link. This assumption   allows the representation  of PDF and CDF  in terms of  incomplete Gamma functions.  
\begin{my_corollary} \label{cor:pdf_single}
	The PDF and CDF of SNR with the effect of random path loss and antenna misalignment errors with transceiver hardware impairments for the THz link are given by
	\begin{flalign} \label{eq:pdf_hl_hp}
	&f_{\gamma}(\gamma) =  -\frac{z^k \rho^2 a_l^{z-\rho} \gamma^{\frac{\rho-1}{2}} (z-\rho)^{-k}}{\Gamma(k) \bar{\gamma_h}^{\frac{\rho-1}{2}}} \frac{1}{2\big(1-\gamma k_h^2\big)\sqrt{\frac{\bar{\gamma} \gamma}{1-\gamma k_h^2}}} \nonumber \\ & \Bigg[ \frac{1}{z-\rho} \Big[\Gamma(k+1) - \Gamma\Big(k+1, (z-\rho) \ln \Big(\frac{a_l \bar{\gamma}}{\gamma_h}\Big)\Big)\Big]  \nonumber \\ &  - \Big[\Gamma(k) - \Gamma\Big(k,(z-\rho) \ln \Big(\frac{a_l \bar{\gamma}}{\gamma_h}\Big)\Big)\Big] \Bigg]
	\end{flalign}	
	
	\begin{flalign} \label{eq:cdf_hl_hp}
	& F_{\gamma}(\gamma) = - {z^k \rho^2 a_l^{z} (z-\rho)^{-k}} \Bigg[ \frac{k}{z-\rho} \bigg[\frac{1}{\rho} \Big(\ln\Big(\frac{a_l \bar{\gamma}}{\gamma_h}\Big)\Big)^{-\rho} \nonumber \\ & + \sum_{j=0}^{k} \frac{\Big((z-\rho)\Big)^j}{j!} (z+2\rho)^{-j-1} \Gamma\Big(j+1,(z+2\rho) \ln\Big(\frac{a_l \bar{\gamma}}{\gamma_h}\Big)\Big) \bigg] \nonumber \\ & + \bigg[\frac{1}{\rho} \Big(\ln\Big(\frac{a_l \bar{\gamma}}{\gamma_h}\Big)\Big)^{-\rho} + \sum_{j=0}^{k-1} \frac{\Big((z-\rho)\Big)^j}{j!}(z+2\rho)^{-j-1} \nonumber \\ & \times \Gamma\Big(j+1,(z+2\rho) \ln\Big(\frac{a_l \bar{\gamma}}{\gamma_h}\Big)\Big)\bigg] \Bigg]
	\end{flalign}
\end{my_corollary}
\begin{IEEEproof}
	The proof is presented in Appendix B.
\end{IEEEproof}
By utilizing the statistical findings presented in Theorem 1, Corollary 1, and Corollary 2, we can derive analytical expressions for various performance metrics in THz wireless systems.  Moreover, the proposed statistical model  facilitates the development of channel-aware random access protocols, as described in the following section.

\section{Random Access Protocols for multiuser THz transmission} \label{sec:protocol}

This section introduces channel-aware random access protocols to optimize communication in a THz multi-user network. 
The randomness is present at low frequencies mainly due to the shadowing effect or short-term fading. On the other hand, at higher frequency bands, various factors could cause randomness in the propagation environment. In the THz link, we consider the random effects of path loss, short-term fading, and antenna misalignment errors, along with the effects of transceiver hardware impairment.  Since there are so many random effects in the THz link, a number of users might have a signal strength lower than the predefined quality of service (QoS). As a result, the number of users accessing the channel becomes limited. Thus, we capitalize on the inherent variability in channel conditions to formulate a random access protocol specifically tailored for multi-user THz transmission.

\subsection{Description of the Protocol}
In our approach, we adopt a random access protocol that enables users to access the multi-access channel, without the need for centralized control. Our model assigns each user a single data packet for transmission. The chosen random access scheme is based on the well-known ALOHA protocol due to its simplicity of implementation and straightforward operation. However, we assume that each user is designated a single packet for transmission during   a data frame, necessitating a novel approach to the analysis in contrast to the conventional approach of the slotted ALOHA protocol \cite{rom90book}. Here, a packet refers to the data from a single active user, while a frame denotes the aggregation of data from all active users.

Prior to  data transmission, the protocol necessitates knowledge of the number of active users $K$ who are participating in the process. This information is typically available for the operation and maintenance of the network, and therefore does not add any extra complexity. AP assigns a unique identity (ID) to each active user.

We adopt a collision channel model, where a successful transmission is only guaranteed when a single user transmits at a given time. In this case, the AP receives a unique packet and sends an "ACK" message  to the successful user. Once acknowledged, the user clears its buffer and remains non-operative until the transmission of the current frame is completed. However, if more than one user transmits simultaneously, a collision occurs. In such cases, the users involved in the collision wait for a another slot and then resume their transmission attempts until successful transmission conditions are met. There may be instances when no user transmits at a given time, resulting in an idle multi-access channel for that duration. Thus,  channel idle time increases the latency of data transmission, while collisions increase both latency and energy consumption for users.

The occurrence of successful transmission, which primarily depends on the TP of each user, is highly desirable. We propose two schemes: the FTP scheme and the ATP scheme. The FTP scheme offers a simple procedure for assigning the TP, while the ATP scheme aims to achieve optimal performance in each transmission. For a given network of $K$ transmitting users, a stable transmission is ensured if the probability of transmission for each user is set as $p = \frac{1}{K}$ \cite{Abramson1970}. This probability of transmission is assigned to each user at the beginning of the data collection process in both schemes. In the FTP scheme, once assigned, the probability of transmission remains constant for the entire duration of the frame. On the other hand, in the ATP scheme, the TP is dynamically adapted with each successful transmission of a packet, utilizing ACK/NACK messages, to ensure stable transmission for each packet during the frame. The algorithm is described in Algorithm \ref{euclid}.

In what follows next, we evaluate delay and energy consumption using FTP and ATP protocols.  We also analyze outage performance following successful transmission, utilizing the derived generalized channel model from the previous section.

\begin{algorithm}[t]
	\caption{FTP and ATP Protocols }\label{euclid}
	\begin{algorithmic}[1]
		\Procedure{Data transmission for frame $i$}{}
		\State AP performs network discovery to compute number of users $N$ with a data packet. 
		\State AP considers a threshold SNR $\gamma_{\rm QoS}$ based on quality of service.
		
		\State AP estimates SNR $\gamma_i$, $i=1, 2, \cdots, N$ from each user.
		\State AP selects $K$ active users only if   the SNR $\gamma_i>\gamma_{\rm QoS}$, $i=1, 2, \cdots, N$. The THz propagation may ensure $K\ll N$.
		\BState \emph{loop 1}
		\State AP assigns a transmission probability to each user $p=\frac{1}{K}$
		\BState \emph{loop 2}
		\State All $K$ active users transmits with probability $p$
		\If {  multiple users transmit} AP \Return "NACK" failure messages: \textbf{goto} \emph{loop 2}  (for both FTP and ATP)
		\Else~ AP \Return "ACK" success message and $K \gets K-1$: \textbf{goto} \emph{loop 1}(ATP), \textbf{goto} \emph{loop 2} (FTP)
		\EndIf
		\If {  $K=0$} $i\gets i+1$
		\EndIf
		\State \textbf{goto} \emph{top}.
		\EndProcedure
	\end{algorithmic}
\end{algorithm}

\subsection{Performance Analysis for Delay and Scaling Laws}
 Taking into account the probabilities of multiple transmissions (collisions) and no transmission (idle channel), we can express the expected delay for a successful transmission as $\mathbb{E} [{D_k}]= \sum_{k=1}^{\infty}kP_s(1-P_s)^{k-1} =\frac{1}{P_s}$, where $P_s = kp(1-p)^{(k-1)}$ represents the probability of successful transmission in a network of $k$ users. Consequently, the expected $D_K$ required to successfully collect data from $K$ users can be expressed as:
\begin{align}
D_K=\sum\limits_{k=1}^K \frac{1}{kp(1-p)^{(k-1)}}.
\label{eq:M_K}
\end{align}
The equation (\ref{eq:M_K}) demonstrates that the number of transmissions is predominantly influenced by the number of active users, denoted as $K$, and their respective transmission probabilities, represented by $p$. 

In the FTP scheme, we  substitute $p = \frac{1}{K}$ in (\ref{eq:M_K}) to compute the number of transmissions required by the FTP scheme.  We denote by $r=1-\frac{1}{K}<1$,  and  $r^{k-1} = \exp\big({(k-1)\log r}\big)$ to simplify  (\ref{eq:M_K}):
\begin{align}
D_{\rm{FTP}} =  (K-1)\sum\limits_{k=1}^K \frac{1}{k\cdot \exp(k\log r)}.
\label{eq:M_FTP}
\end{align}
\begin{my_lemma}
	The expected number of transmissions required by the random access FTP scheme  to transmit data from $K$ active users can be bounded by
	\begin{align}
	\label{lemma:ftp_bound}
	\begin{split}
	(K-1)\Big(\log K+\frac{1}{1+2K}+\gamma+1\Big)<	D_{\rm{FTP}}
	\\
	< (K-1)\Big(\log K+\frac{1}{K(K-1)}+\frac{K}{K-1}\mathrm{e}+1\Big),
	\end{split}
	\end{align}
	and thus scales as		
	$M_{\rm{FTP}} =	\mathcal{O}(K\log K)$.
\end{my_lemma}

\begin{IEEEproof}
The proof is presented in Appendix C.
\end{IEEEproof}

When the number of active users decreases, the constant transmission probability in the FTP scheme results in a reduction in collisions but an increase in delay for subsequent successful transmissions. This delay is primarily due to longer periods of channel idle time without any transmissions. While this increase in delay has a minimal impact on energy consumption for users, it introduces latency in data collection. To address this latency issue, the ATP scheme can be implemented.

The ATP scheme is  more stable as it continually updates the transmission probability with each successful transmission. Thus, we substitute $p = 1/k$ in \ref{eq:M_K}) to obtain:
\begin{align}\label{eq:M_ATP}
D_{\rm{ATP}}=\sum\limits_{k=1}^K \left(\frac{k}{k-1}\right) ^{(k-1)}.
\end{align}

\begin{my_lemma}
	The expected number of transmissions required by the random access ATP scheme  to transmit data from $K$ active users can be bounded by
	\begin{align}
	\label{lemma:atp_bound}
	K\mathrm{e}-\mathrm{e}(\gamma+\log K +\frac{1}{2K})\leq D_{\rm{ATP}}\leq K\mathrm{e},
	\end{align}
	and thus scales as 	$D_{\rm{ATP}} =	 \mathcal{O}(K\mathrm{e})$.	 	
\end{my_lemma}
\begin{IEEEproof}
	The expression (\ref{eq:M_ATP}) can be rewritten as $D_{\text{ATP}} = \sum\limits_{k=1}^K \exp\left((k-1)\log\left(1+\frac{1}{k-1}\right)\right)$. By applying the logarithm inequality $\log\left(1+\frac{1}{k-1}\right) < \frac{1}{k-1}$, we establish the upper bound stated in (\ref{lemma:atp_bound}).
	
	For the lower bound, we can utilize the inequalities $\log\left(1+\frac{1}{k-1}\right) > \frac{1}{k}$ and $\exp\left(-\frac{1}{k}\right) > 1-\frac{1}{k}$ to obtain $D_{\text{ATP}} > \mathrm{e}K - \mathrm{e} \mathcal{H}_K$. By using an upper bound on the harmonic number, $\mathcal{H}_K < \gamma + \log K + \frac{1}{2K}$ \cite{Abramowitz1972book}, we derive the lower bound as stated in (\ref{lemma:atp_bound}). Furthermore, it is evident that the scaling law holds true as $K \rightarrow \infty$ in (\ref{lemma:atp_bound}).
\end{IEEEproof}

It is worth highlighting that the ATP scheme requires only approximately $\mathrm{e}$ times more transmissions compared to optimal scheduling, where centralized control is necessary to enable sequential transmissions from selected users.

Finally, we utilize Hoeffding's concentration inequality to illustrate the tightness of the proposed average delay analysis. This is achieved by examining the deviation (represented by $\epsilon$) of the probability of the sample delay  from all users: 
\begin{flalign}
	{\rm Pr}[|\bar{D}_K-D_K|>\epsilon]\leq 2\exp\bigg(-\frac{2\epsilon^2}{n }\bigg)\leq 2\exp\bigg(-\frac{2\epsilon^2}{n}\bigg)
\end{flalign}
where $ \bar{D}_K $ is the sample delay (obtained with $n$ samples) and $ D_K $ is the average delay. It can be seen that the sample delay can provide higher accuracy for a moderate value of $n$ for real-time simulations.

\begin{figure*}[tp]
	\subfigure[PDF of channel gain $|h|$.]{\includegraphics[scale=0.27]{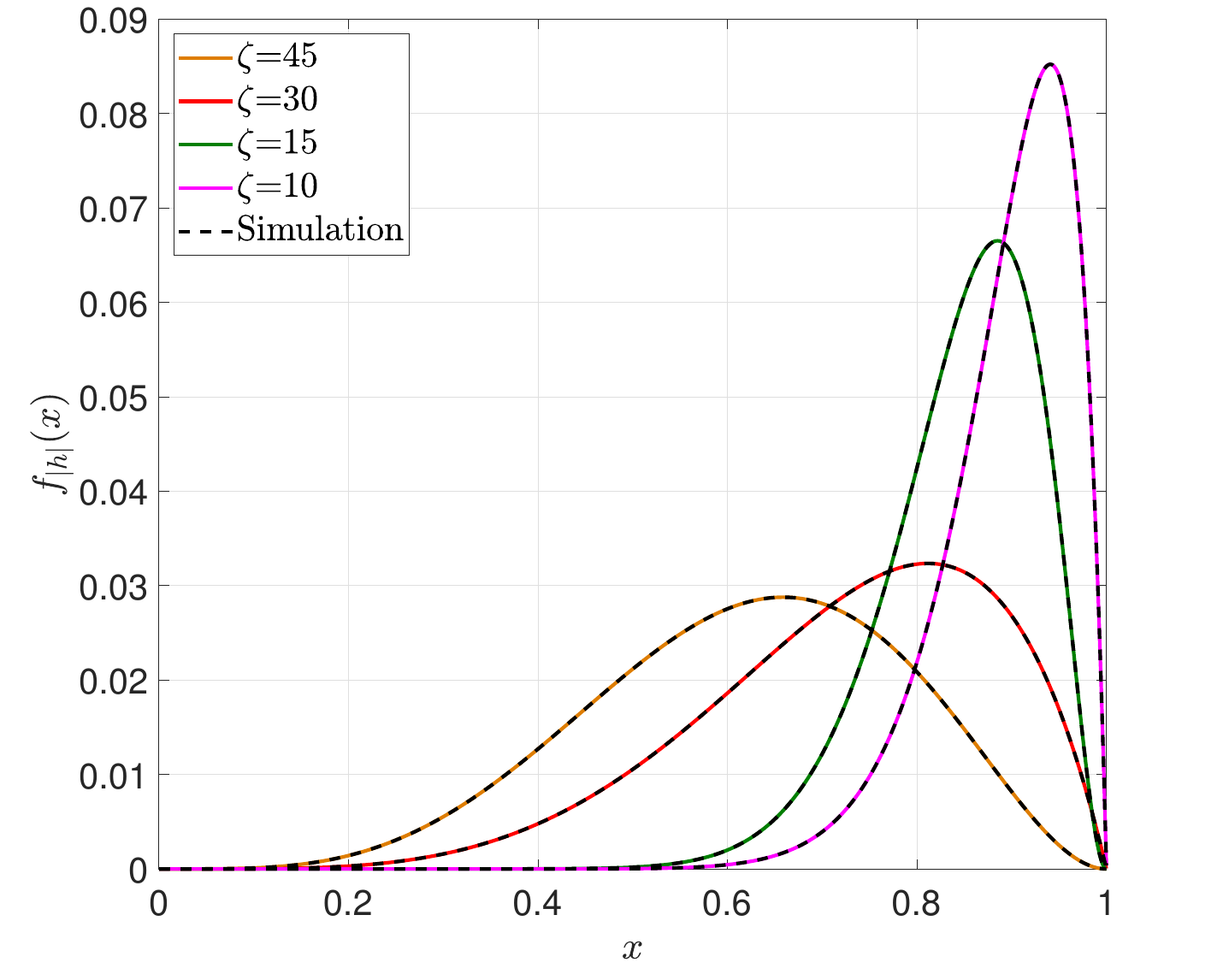}}\hspace{-6mm}
	\subfigure[CDF of channel gain $|h|$.]{\includegraphics[scale=0.27]{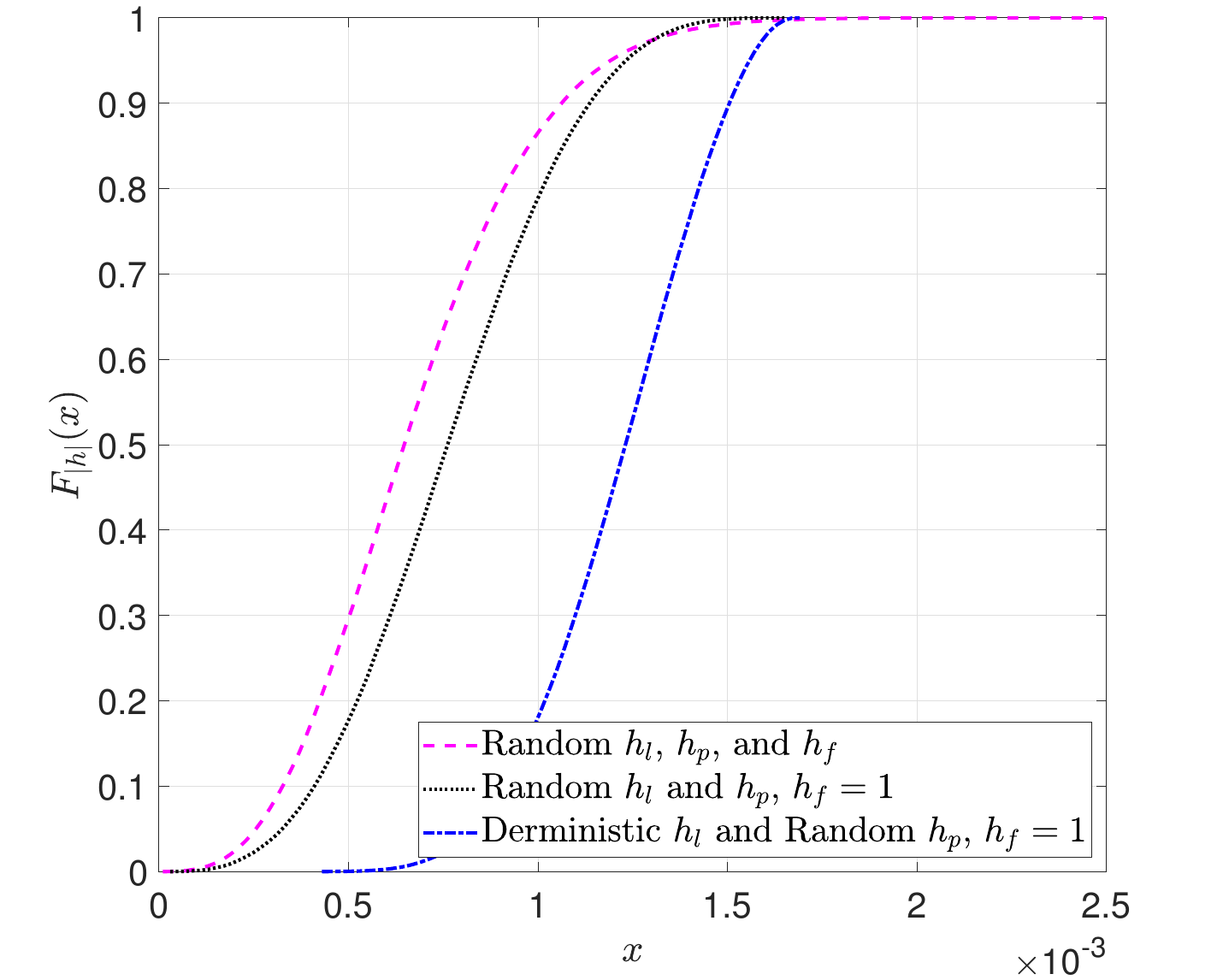}}
	\subfigure[Outage probability with different $\zeta$  at  $d$=$100$ \mbox{m}.]{\includegraphics[scale=0.27]{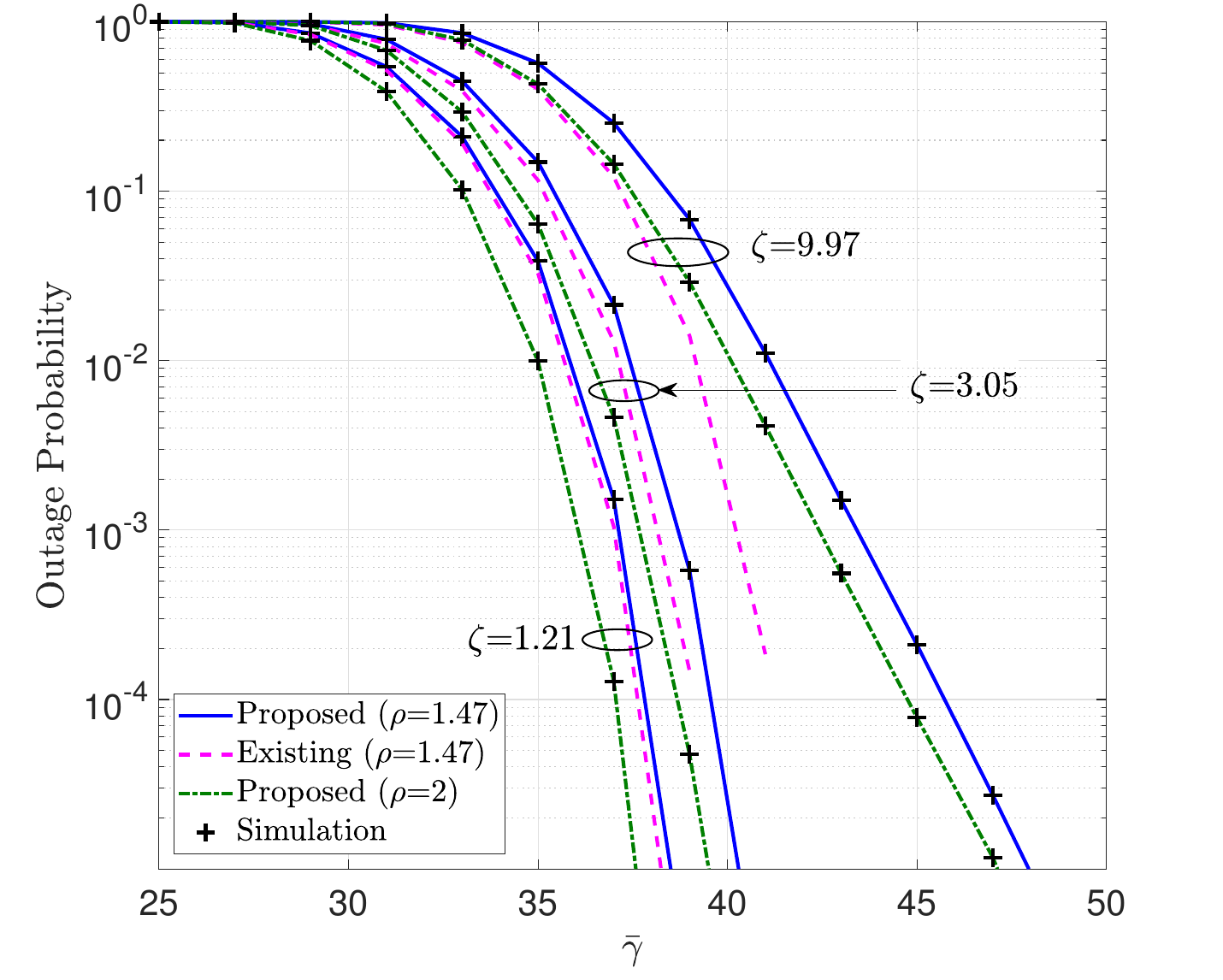}}
	\caption{Generalized channel model for THz propagation with $\alpha=2$, $\mu=1$, $\eta=1$, $\kappa=1$, $\rho=4$.}
	\label{fig:pdf}
	
\end{figure*}

\subsection{Analysis for Energy Consumption Performance}
In this subsection, we examine the expected energy required to collect the data as a parameter to analyze the  energy consumption of the  protocol.  We derive  analytical bounds on the expected energy of both ATP and FTP  with scaling laws.

In the $p$-persistent slotted ALOHA protocol each node transmits with probability $p$. Therefore, when we have $m$ users that have data to transmit, the distribution of the number of users accessing the channel is binomial: $P[Z=m]= {{k}\choose{m}}p^m(1-p)^{k-m}$, where $k\leq K$ and $K$ is the total number of active users. Hence the expected number of colliding packets in each transmission given that the transmission was unsuccessful is denoted by $\mathbb{E}[N_{c}]$ (i.e. number of collisions per transmission event given that it failed):
\begin{align}
	\begin{split}
		\mathbb{E}[N_{c}]= \sum_{m=2}^k mP[Z=m]=kp-kp(1-p)^{k-1}.
	\end{split}
	\label{eq_Ntx}
\end{align}
The expected number of transmission attempts  between two successful consecutive  data transmissions  is given by
\begin{align}
	\begin{split}
		\mathbb{E}[N_{\rm tx}]=\sum_{k=1}^{\infty}kP_s(1-P_s)^{k-1} =1/{P_s},
		\label{eq_Dk}
	\end{split}
\end{align}
where  $ P_s=kp(1-p)^{(k-1)}$ is the probability of success.
Using (\ref{eq_Ntx}) and (\ref{eq_Dk}), the expected energy consumed until a single successful transmission of a single packet out of $k$ users that still need to transmit is given by:
\begin{align}
	\begin{split}
		\mathbb{E}[E_{k}] =\mathbb{E}[N_{tx}] \mathbb{E}[N_{c}] E_{\rm tx}+E_{\rm tx}= \frac{E_{\rm tx}}{(1-p)^{(k-1)}},
	\end{split}
\end{align}
where plus $E_{\rm tx}$ corresponds to the final successful user transmission. The constant $E_{\rm tx}$ represents the energy consumption per single packet transmission. We assume $E_{\rm tx}=1$ for the following analysis.
The expected energy consumed to collect data from $K$ users is given by:
\begin{align}
	\begin{split}
	E_{K}=	\mathbb{E}[E_{k}]= \sum_{k=1}^{K}\Big[\frac{1}{(1-p)^{(k-1)}}\Big].
	\end{split}
	\label{eq_EK}
\end{align}	
In what follows, we derive performance bounds on the consumed energy by the FTP and ATP schemes.

We  substitute $p = {1}/{K}$ in (\ref{eq_EK}) to compute the expected energy $E_{\rm{FTP}}$ required by the FTP scheme. We denote by $r=1-{1}/{K}<1$,  and  $r^{k-1} = \exp\big({(k-1)\log r}\big)$ to simplify  (\ref{eq_EK}):
\begin{eqnarray}
	E_{\rm{FTP}}=\frac{K-1}{K}\sum_{k=1}^{K}\Big[\frac{1}{\exp\big({k\log r}\big)}\Big].
	\label{eq_E_FTP}
\end{eqnarray}	
\begin{my_lemma}
	The expected energy consumption  by the random access FTP scheme with $K$ active users can be bounded by
	\begin{align}
		\label{lemma:ftp_Ebound}
		\begin{split}
			\frac{3K}{2}-\frac{1}{2K}-1<	\eta_{\rm{FTP}}
			<	\frac{\rm{e}(K-1)^2}{K(K-2)}-1+1/K.
		\end{split}
	\end{align}
	and thus does not scale with the number of 	users.	
\end{my_lemma}

\begin{IEEEproof}
	Applying upper bound $\log r <-{1}/{K}$ and then lower bound $\exp(k/K) >1+{{k}/{K}}$ in  (\ref{eq_E_FTP}), we get the lower bound in (\ref{lemma:ftp_Ebound}). For the upper bound, we use geometric series to sum  (\ref{eq_E_FTP}) as $(K-1)[\exp(-K\log(r))-1]$. Using  $\log(1-1/K)>1/(1-K)$ and $\exp(1/(K-1))<(K-1)/(K-2)$, the expression is simplified as ${(K-1)}[\rm{e}(K-1)/(K-2)-1]$ and thus we get the upper bound in (\ref{lemma:ftp_Ebound}).		
	The lower and upper bounds in  (\ref{lemma:ftp_bound}) reveal that the scaling law   holds true as  $K\rightarrow \infty$.
\end{IEEEproof}

Since the transmission probability is held constant even with a decrease in the number of active users, there is a decrease in the number of collisions but an increase in the delay for subsequent successful transmissions.
This delay is due to the increase in the channel idle time with no transmissions, and thus has a negligible effect on the energy consumption of the users.   It is emphasized that the expected energy consumption in the  FTP scheme is merely  $\mathrm{e}-1$ times more than the  optimal scheduling where a centralized control is required to allow sequential transmissions from the selected users. 

To compute the expected energy with the ATP scheme, we  substitute  $p = {1}/{k}$ in (\ref{eq_EK}) to get
\begin{align}\label{eq:E_ATP2}
	E_{\rm{ATP}}=\sum\limits_{k=1}^K \left(\frac{k}{k-1}\right) ^{(k-1)}.
\end{align}
\begin{my_lemma}
	The expected 	energy consumption  by the random access ATP scheme  to collect data from $K$ active users can be bounded by
	\begin{align}
		\label{lemma:atp_Ebound}
		\begin{split}
			\mathrm{e}-\frac{\mathrm{e}}{K}(\gamma+\log K +\frac{1}{2K})\leq E_{\rm{ATP}}\leq \mathrm{e},
		\end{split}
	\end{align}
	and thus does not scale with number of users.		
\end{my_lemma}	 	
\begin{IEEEproof}	
	We can represent  (\ref{eq:E_ATP2}) as	$E_{\rm{ATP}}= \sum\limits_{k=1}^K \exp\big({(k-1)\log(1+{1}/{(k-1)})}\big)$. By  applying the logarithm inequality  $\log\big(1+{1}/{(k-1)}\big)< {1}/{(k-1)}$, we prove the upper bound in	(\ref{lemma:atp_Ebound}).  For the lower bound, we can use the inequalities  $\log\big(1+{1}/{(k-1)}\big) >{1}/{k}$ and $\exp({-{1}/{k}}) > 1-{1}/{k}$ to get 	$E_{\rm{ATP}}>\mathrm{e}K-\mathrm{e} {\cal{H}}_K$, where ${\cal{H}}_K$ is $K$-th harmonic number. Using an upper bound on the harmonic number ${\cal{H}}_K$$ <\gamma+\log K +{1}/{2K}$ \cite{Abramowitz1972book}, we get the lower bound in 		(\ref{lemma:atp_Ebound}).  Further, it can be seen that the scaling law holds true as  $K\rightarrow \infty$  in (\ref{lemma:atp_Ebound}).
\end{IEEEproof}
Using (\ref{lemma:ftp_Ebound}) and (\ref{lemma:atp_Ebound}),  the energy difference $\Delta E_k$ between the ATP and FTP can be bounded as
\begin{align}\label{eq_delta}
	\begin{split}
		(\rm{e}-1){\cal{H}}_K+K-\frac{\rm{e}(K-1)^2}{(K-2)}-1<\Delta E_k\\
		<K \rm e-\frac{3K}{2}+\frac{1}{2K}+1.
	\end{split}
\end{align}

It is important to note that the expected energy consumption of the  ATP scheme is larger than the FTP scheme. This can be easily verified by analyzing the  energy function in (\ref{eq_E_FTP}) and (\ref{eq:E_ATP2}). This function increases monotonically with $p$ and thus the lower transmission probability in the FTP requires less expected energy consumption. Essentially, the ATP scheme consumes $\rm e$ times more than the optimal scheduling whereas the energy consumption in the FTP scheme is $\rm e-1$ times more than the same.

Similar to average delay, we utilize Hoeffding's concentration inequality to illustrate the tightness of the proposed average energy analysis:
\begin{flalign}
	{\rm Pr}[|\bar{E}_K-E_K|>\epsilon]\leq 2\exp\bigg(\hspace{-2mm}-\frac{2\epsilon^2}{n (n-1)^2} \hspace{-1mm}\bigg)
\end{flalign}
where $ \bar{E}_K $ is the sample value and $ E_K $ is the average value.

\subsection{Outage Performance for ATP and FTP Protocols }
In this section, we analyze the outage probability for both ATP and FTP Protocols in the event of successful transmission. In this case, the analysis remains the same for the protocols.

The outage probability of a THz system is the probability that the wireless link quality falls below a specified threshold of SNR $\gamma_{\rm th}$. The outage probability is an important performance metric, which can be obtained using the CDF function $F_\gamma(\gamma)=P(\gamma < \gamma_{\rm th}) $. We can substitute $\gamma = \gamma_{\rm th}$ in  \eqref{eq:cdf_aekm_rpl}, \eqref{eq:cdf_hfpl}, and \eqref{eq:cdf_hl_hp} to get the outage probability for the THz transmission link under different propagation scenarios.

An asymptotic expression for the outage probability can be obtained by invoking $\bar{\gamma}\to \infty$  \eqref{eq:cdf_aekm_rpl}, \eqref{eq:cdf_hfpl}, and \eqref{eq:cdf_hl_hp}. As an illustration, the asymptotic outage probability in high SNR region for \eqref{eq:cdf_aekm_rpl} can be derived using the method of residues as described in \cite{AboRahama_2018}, while the asymptotic outage probability for \eqref{eq:cdf_hfpl} can be obtained using the approach described in \cite[Th. 1.11]{Kilbas_2004}. A general expression for the asymptotic outage probability for both the cases is given by
\begin{flalign} \label{eq:outage_fpl_asymp}
	&P_{\rm out}^\infty(\gamma_{\rm th}) = C_1 \bigg(\frac{\gamma_{h_{\rm th}}}{\bar{\gamma}}\bigg)^{\frac{\alpha\mu}{2}} + C_2 \bigg(\frac{\gamma_{h_{\rm th}}}{\bar{\gamma}}\bigg)^{\frac{\rho}{2}} + C_3 \bigg(\frac{\gamma_{h_{\rm th}}}{\bar{\gamma}}\bigg)^{\frac{z}{2}}
\end{flalign}
where $ C_1 $, $ C_2 $, and $ C_3 $ are constants. Analyzing the exponents of the average SNR in \eqref{eq:outage_fpl_asymp}, the diversity order of the considered system can be obtained as
\begin{flalign}
	&DO = \Biggl\{\frac{\alpha\mu}{2}, \frac{\rho}{2}, \frac{z}{2}\Biggr\}.
\end{flalign}
The diversity order offers multiple options to mitigate the effects of antenna misalignment errors and atmospheric absorption. By understanding the diversity order, we can establish guidelines for effectively utilizing the beam width and link distance to counteract the impact of antenna misalignment errors and random atmospheric absorption. Consequently, the appropriate selection of beam width (to address pointing errors) and link distance (to deal with atmospheric absorption) can help overcome the signal fading.

\section{Simulation and Numerical Analysis} \label{sec:simulation}
In this section, we employ Monte Carlo simulations and numerical computations to illustrate the generalized channel model for THz transmission, considering random path loss. We also assess the effectiveness of suggested random access protocols in a cell-free network by leveraging the statistical characterization of THz propagation.
\begin{figure*}[tp]
	
	\subfigure[Expected number of transmissions ($D_K$).]{\includegraphics[scale=0.26]{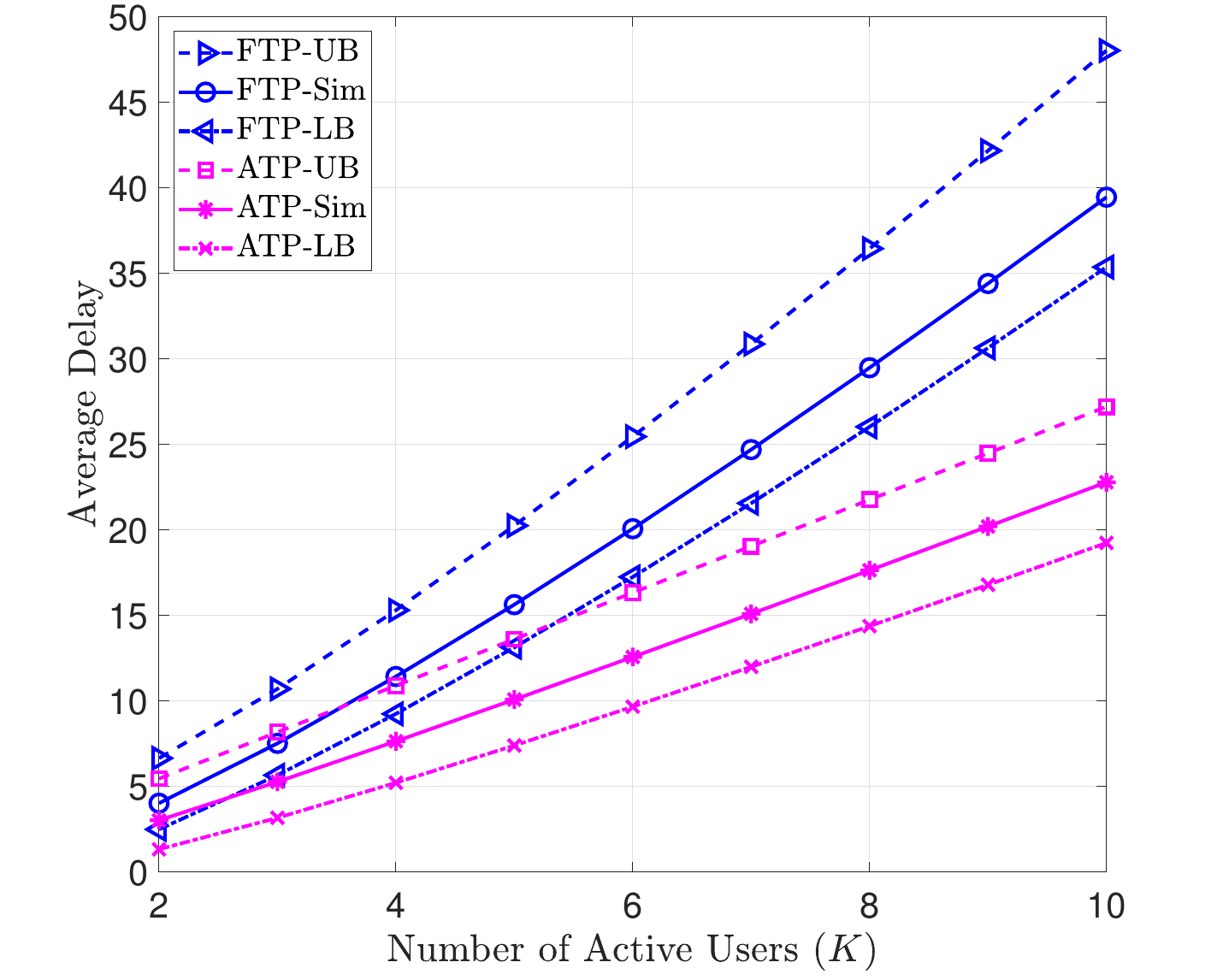}}\hspace{-5mm}
	\subfigure[Energy consumption ($E_K$).]{\includegraphics[scale=0.33]{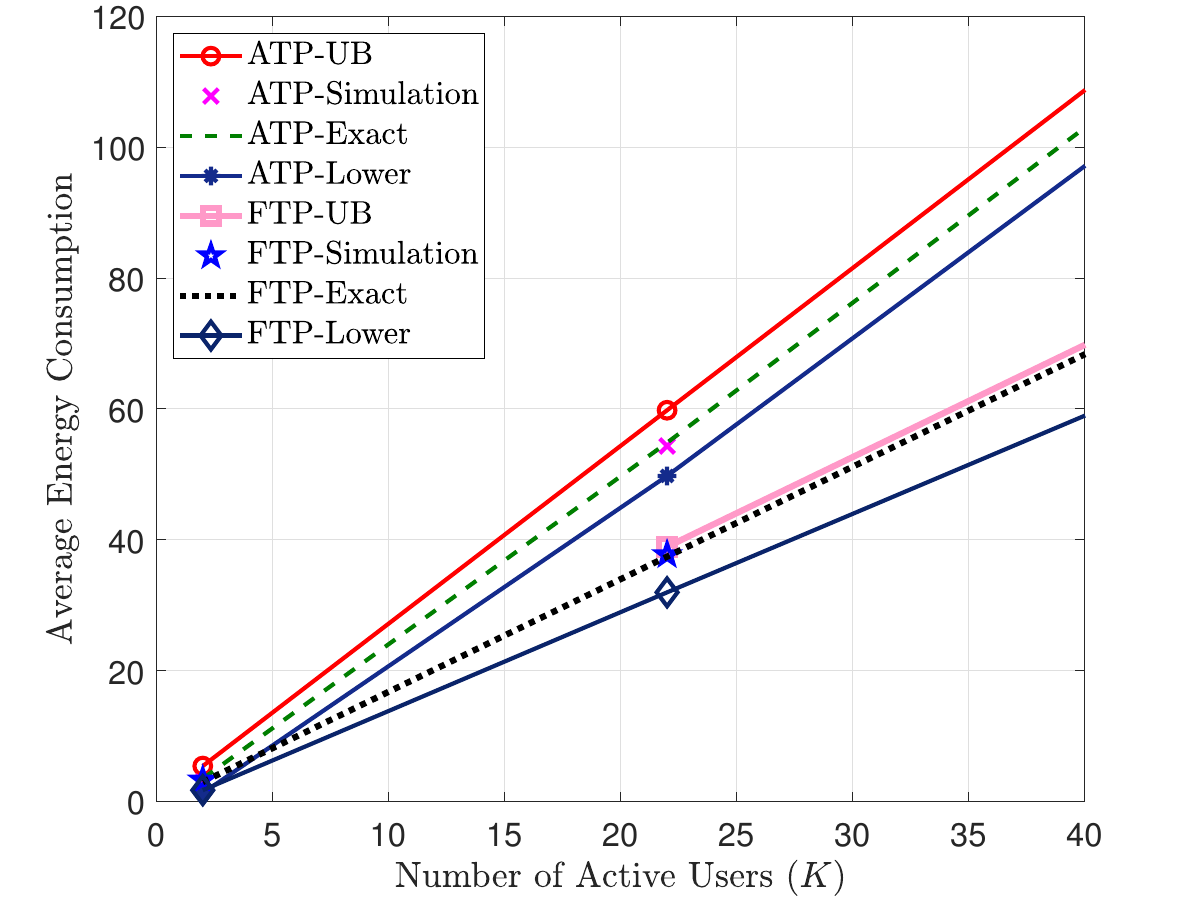}}
	\subfigure[Energy consumption ($E_K$) using realistic parameters. ]{\includegraphics[scale=0.33]{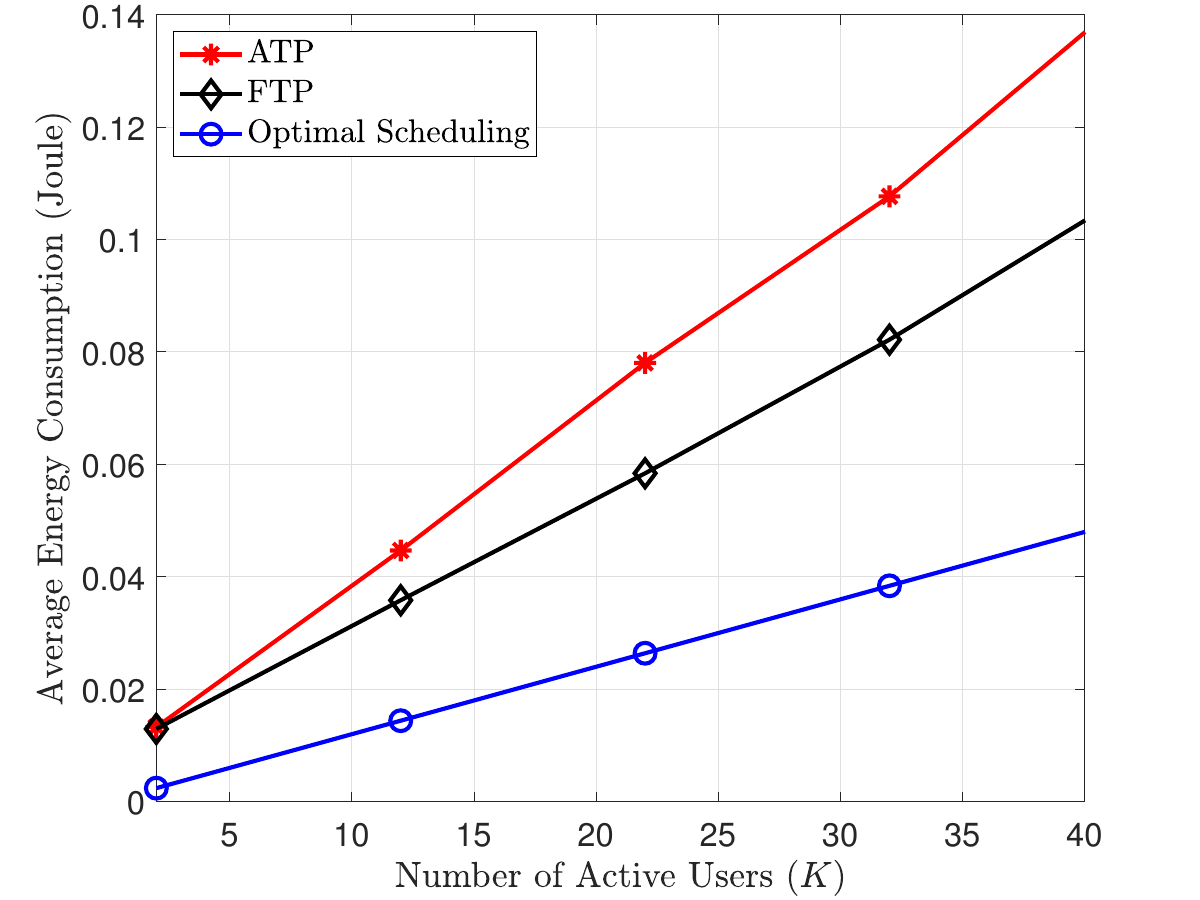}}
	\caption{Delay and energy consumption performance of random access protocols.}
	\label{fig:delay}
\end{figure*}

\subsection{Generalized Channel Model for THz Transmission with Random Path Loss}
In this subsection, we assess the accuracy of the generalized channel model compared with the existing model through simulations and numerical analysis conducted using MATLAB. We utilize Monte Carlo simulations to validate our statistical analysis in random path loss. Our study investigates the THz link performance for numerous channel conditions by adopting different absorption coefficient values $\zeta$. The product of shape and scale parameters for the random path-loss ($k  \beta$) in \eqref{eq:pdf_path_loss} provides the average value of the absorption coefficient $\zeta$ \cite{fog}.

In Fig. \ref{fig:pdf}(a), we illustrate the  PDF of generalized model, as derived in  \eqref{eq:pdf_path_loss} across a wide range of $\zeta$ values, representing different propagation scenarios by considering the impact of random path-loss and antenna misalignment errors. Observing the plots, it is evident that for smaller absorption coefficient values ($\zeta$), the random variable tends to cluster around the value of $1$. Conversely, with higher $\zeta$ values, the variable attenuates and shows a greater tendency to be situated in regions with lower amplitudes. In Fig. \ref{fig:pdf}(b), we simulate the generalized THz statistical model under the combined effect of random atmospheric absorption, non-linearity of fading, hardware impairments, and antenna misalignment errors. The figure shows that the probability of getting a higher threshold value of channel gain decreases with an increase in random characteristics of the channel, motivating for the development of random access protocols over THz band. 

Further, we utilize the simulation environment to demonstrate and verify the effect of the derived random path-loss on the outage probability of the THz link. In Fig.~\ref{fig:pdf}(c), we analyze the outage probability by considering random path-loss and antenna misalignment errors. For this analysis, we exclude the effect of short-term fading and transceiver hardware impairments on the THz link's performance. The plots clearly demonstrate that with increases in $\zeta$ value, the outage probability increases. Further, we demonstrate the impact of antenna misalignment errors on the outage probability performance. The system's outage performance improves as the antenna misalignment error parameter $\rho$ increases. We compare our proposed results with existing ones employing a deterministic $\zeta$. Figure \ref{fig:pdf}(c) clearly shows that using the deterministic model of $\zeta$ sets an upper bound on the performance of the THz link. At an SNR of $ 45 $ \mbox{dB}, the outage probability is nearly four times lower when employing the random path-loss model compared to the deterministic path-loss model used in previous works. This highlights the importance of considering the random path-loss model to accurately capture the behavior of the actual system.

\begin{figure*}[tp]
	\begin{center}	
		\subfigure[Channel parametrs $k=1$ and $\alpha=1$ for various $\rho$ and $\mu$. ]{\includegraphics[scale=0.30]{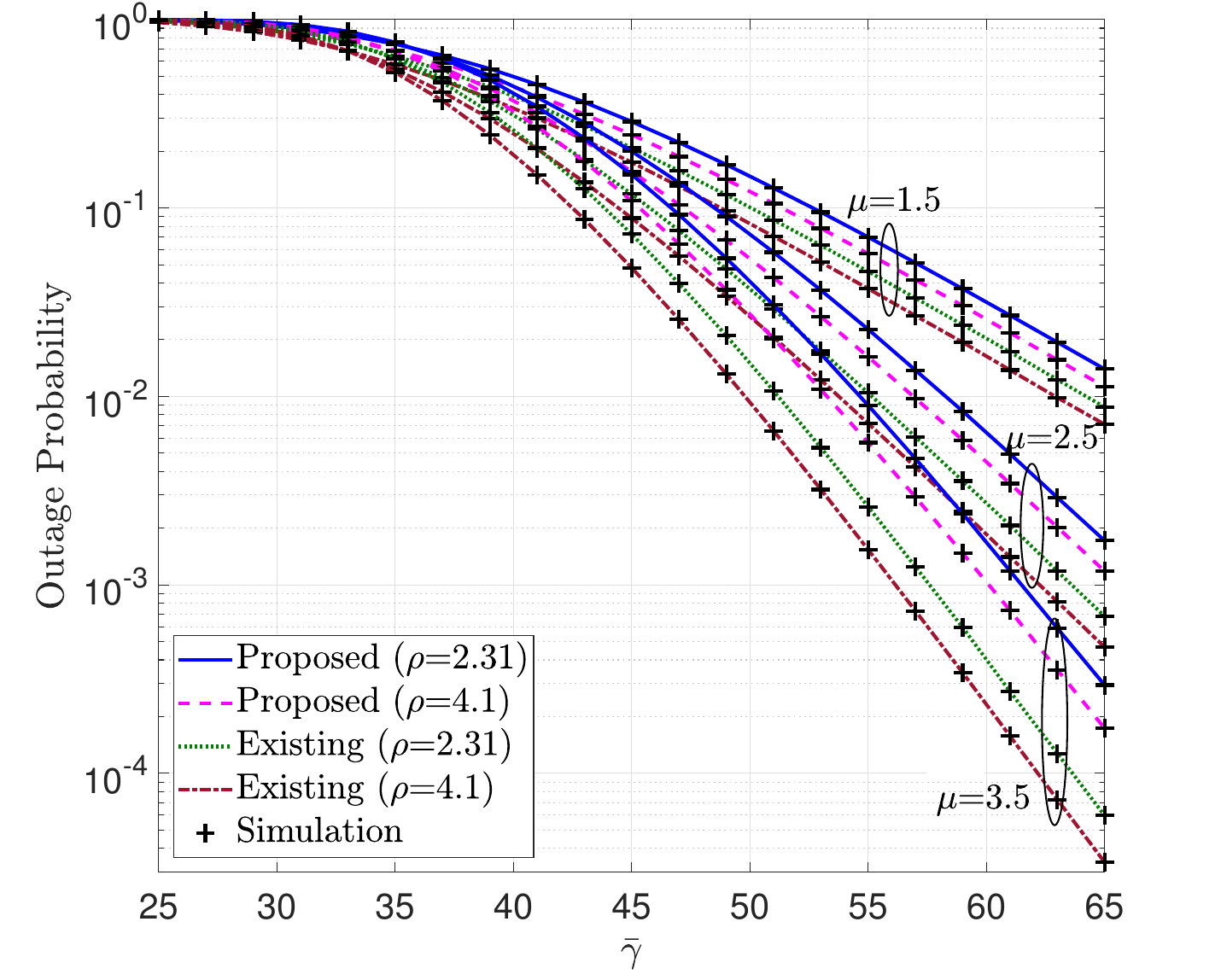}}\hspace{4mm}
		\subfigure[Channel paramerters  with $k=3$, $\mu=2$, $\eta=1$, and $\rho=4.1$ for  various $\alpha$, $ \kappa $, and $ k_h $. ]{\includegraphics[scale=0.30]{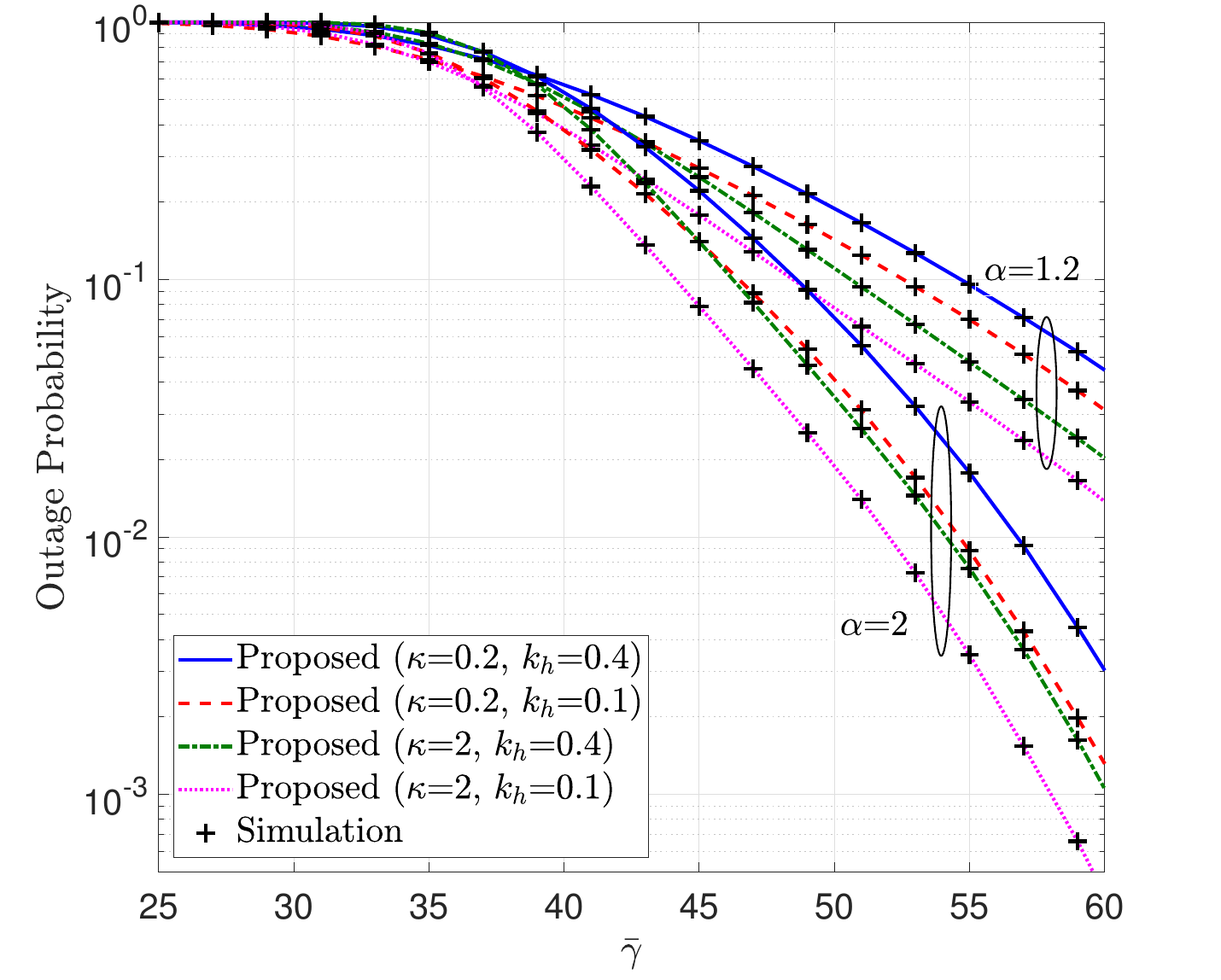}}
		\caption{Outage probability performance for FTP/ATP protocol over generalized THz channel model at $d=100$ \mbox{m}.}
		\label{fig:outage_hfpl}
	\end{center}
\end{figure*}

\subsection{Random Access Protocols for Cell-Free Network}
In this section we analyze the efficacy of the proposed random access protocols with respect to delay and energy consumption in a multiuser environment. We simulate  random access protocols  transmitting randomly with  assigned transmission probabilities based on the FTP and ATP schemes. The simulation results of the number of transmissions (which also includes multiple transmissions due to the collision) required for each successful transmission were averaged over  $5000$ trials. For delay analysis, we assign unit time for each unsuccessful data collection event, which includes either the collision or the idle state each consuming a single unit time. For energy consumption, first, we  assign unit energy for each transmission from users (including number of collisions) and  neglect the energy consumed during the reception of ACK messages and in the event of idle condition of the users. Later, we use actual energy consumed during data transmission and other overheads.

In Fig.~\ref{fig:delay}, we analyze the expected delay and energy consumption by the proposed scheme by considering a fixed number of active users ($K$ = $1$ to $10$) in a network. 
As shown in Fig.~\ref{fig:delay}(a), the adaptive scheme always performs better and incurs less delay than the FTP scheme for data collection. It can be observed form the Fig.~\ref{fig:delay}(a) that there occurs more delay for the FTP scheme and number of transmission attempts almost doubles for FTP scheme as compared with the ATP scheme for $10$ active users.  Further,  simulation results can be seen to be in excellent agreement with the exact expressions and derived bounds, as depicted in Fig.~\ref{fig:delay}(a).

   Fig.~\ref{fig:delay}(b) shows that the energy consumption is almost $ 1.5 $ times more for ATP scheme at $40$ active users since FTP incurs more idle slots than the ATP. It can be seen that the relative gain in the energy by the ATP scheme approaches to $1/\rm{e}$ with an increase in the number of active users. The plots in Fig.~\ref{fig:delay}(b) show that the an increase in the delay in the FTP scheme has a negligible effect on its energy consumption; in fact the FTP consumes less energy than the ATP scheme due to reduced collisions. We also compare the derived analytical bounds with the simulation and numerical results in Fig.~\ref{fig:delay}(b).  The derived bounds for energy consumption are shown to be an excellent match with the exact results except  the lower bound of the FTP scheme. Further, we demonstrate energy consumption using realistic parameters, which includes energy consumption  during the reception of ACK messages and energy consumed in the idle state of users with a packet of data. We use transmission  energy for data packet   $1200$ $\mu$J, energy for ACK message     $120$ $\mu$J, and idle energy consumption per user $40$ $\mu$J \cite{biswas2015}.  The performance is compared with the optimal scheduling. It is shown in Fig.~\ref{fig:delay}(c) that the simple FTP scheme (without collision control and adaptation of TP during data collection) requires less energy than the ATP scheme, and can be effective to prolong  the life time of network. However, the advantage of reduced energy consumption in the FTP scheme  comes with an increased delay in collecting the data, as demonstrated in  Fig.~\ref{fig:delay}(a). Although the optimal scheduling requires less energy than the proposed random access protocols, it requires centralized processing or more time/frequency resources.

Finally, Fig. \ref{fig:outage_hfpl} demonstrates the outage performance for successful transmission by considering random path-loss, antenna misalignment error, $\alpha$-$\eta$-$\kappa$-$\mu$ short-term fading, and transceiver hardware impairments. In Fig. \ref{fig:outage_hfpl}(a), we present the outage performance of the THz link with the channel fading parameter $\mu$ and the antenna misalignment error parameter $\rho$. The outage probability decreases as the number of multi-path clusters increases. Further, the outage performance improves with an increase in $\rho$, indicating a decrease in antenna misalignment error. It can be observed from the plots that the outage performance improves by approximately $ 30 $ times when $\mu$ is increased from $ 1.5 $ to $ 2.5 $ at an SNR of $ 60 $ \mbox{dB} for $\rho = 4.1$.  Fig. \ref{fig:outage_hfpl}(b) shows that the THz link's outage performance improves with higher values of the fading parameters $\alpha$ and $\kappa$. Further, the impact of transceiver hardware impairment is evident in the outage probability. As the hardware impairment coefficient $k_h$ increases, the outage probability also increases. When $\alpha$ increases from $1.2$ to $2$ for $\kappa=0.2$ and $k_h=0.4$ at an SNR of $55$ dB, the outage probability is reduced by nearly $9$ times. The figure demonstrates that the slope of the plots remains constant when $\rho$ and $ k $ ($z = 8.686/(\beta d)$) values change but changes only with $\alpha$ and $\mu$, thereby confirming our diversity order analysis.

\section{Conclusion}\label{section_conclusion} \label{sec:conc}

In this paper, we developed a generalized THz transmission model, including random path-loss, antenna misalignment errors, short-term fading, and transceiver hardware impairments.   By accounting for the random path loss, generalized short-term fading, and transceiver hardware impairment, the proposed channel model can better capture the actual behavior and performance of the system at  THz frequencies. Leveraging the statistical model, we proposed random access protocols for a cell-free wireless network to ensure transmission for multiple users while minimizing delay and energy consumption. The protocols adapt the transmission probability using a-priori estimates of the number of active users with data to transmit. We derived tight bounds on the performance of both schemes and showed that the simple FTP scheme incurs reduced energy consumption compared to the ATP  at the cost of an increased expected delay. Computer simulations demonstrated the efficacy of the proposed random access schemes and accuracy in performance assessment with the statistical effect of THz propagation for a cell-free network.   Validating the proposed model and developing protocols with experimental data would be a possible avenue for future research.

\section*{Appendix A}
To derive the combined PDF of random path-loss, antenna misalignment error, and short-term fading, we first derive the joint PDF of short-term fading and antenna misalignment error $f_{h_{\rm fp}}(x)$, which is given by \cite{papoulis_2002}
\begin{flalign} \label{eq:combined_hfp_eqn}
f_{h_{fp}}(z) = \int_{z}^{\infty} \frac {1}{x} f_{h_{f}}(x)  f_{h_{p}}\left ({\frac {z}{x}}\right)  \mathrm {d}x.
\end{flalign}

Substituting the respective PDFs of short-term fading and antenna misalignment error from \eqref{eq:pdf_aekm} and \eqref{eq:pdf_pointing_thz} in \eqref{eq:combined_hfp_eqn}, utilizing the Mellin Barnes type integral form of the exponential function and substituting $ \ln\big(\frac{z}{x}\big)=t $, we rewrite the integration and get the inner integral $ \int_{0}^{\infty} {(e^{-t})^{\alpha(1+A_1+A_2+s_1+s_2+s_3+s_4)-\rho+1}}  t \mathrm {d}t $. We solve the inner integral by applying the identity \cite[(3.381/4)]{Gradshteyn} and after some mathematical manipulation and utilizing the definition of multivariate Fox's H-function, we get the joint PDF of short-term fading and antenna misalignment error as
\begin{flalign} \label{eq:pdf_hfp_aekm}
&f_{h_{fp}}(z) = \frac{ \psi_1 \rho^2 (z)^{\alpha(1+A_1+A_2)}}{(\hat{r}^\alpha)^{1+\frac{\mu}{2}}} \nonumber \\ &\times H^{0,2;1,0;1,1;1,0;1,0}_{2,2;0,1;2,3;1,3;0,1} \Bigg[ \begin{matrix}~ V_3~ \\ ~V_4~ \end{matrix} \Bigg| A_{3}z^{\alpha}, \frac{A_4^2}{4}z^{\alpha},\frac{A_5^2}{4}z^{\alpha},\psi_3z^{\alpha} \Bigg],
\end{flalign}
where $V_3 = \big\{(-A_2;1,0,1,0)\big\}, \big\{\rho-\alpha-\alpha A_1 -\alpha A_2; \alpha, \alpha, \alpha, \alpha\big\}: \big\{(-,-)\big\} ; \big\{(-A_1,1)(\frac{1}{2},1)\big\} ;\big\{(\frac{1}{2},1)\big\}; \big\{-, -\big\} $ and $V_4 = \big\{(-1-A_1 -A_2;1,1,1,0) \big\}, \big\{\rho-1-\alpha-\alpha A_1 -\alpha A_2; \alpha, \alpha, \alpha, \alpha\big\} : \big\{(0,1) \big\} ; \big\{(0,1),(-A_1,1),(\frac{1}{2},1) \big\} ; \big\{(0,1),(-A_2,1),(\frac{1}{2},1) \big\} ; \\ \big\{ (0,1)\big\}$.

Now, we need to multiply the joint PDF derived in \eqref{eq:pdf_hfp_aekm}, with the PDF of random path-loss in \eqref{eq:pdf_path_loss} to derive the combined PDF of short-term fading, antenna misalignment error, and random path-loss. The combined PDF equation is given by \cite{papoulis_2002}
\begin{flalign} \label{eq:combined_hfpl_eqn_aekm}
f_{h_{fpl}}(y) = \int_{\frac{y}{a}}^{\infty} \frac {1}{x} f_{h_{fp}}(x) f_{h_{l}}\left ({\frac {y}{x}}\right) \mathrm {d}x.
\end{flalign}

Similarly, plugging the PDFs of \eqref{eq:pdf_path_loss} and \eqref{eq:pdf_hfp_aekm} in \eqref{eq:combined_hfpl_eqn_aekm} and substituting $ \ln\big(\frac{ax}{y}\big) = t $, and following the similar procedure, we get the combined PDF of fading, antenna misalignment error and random path-loss in \eqref{eq:pdf_aekm_rpl}. The PDF of SNR can be derived by simple transformation of random variable \cite{papoulis_2002} as $f_{\gamma}(\gamma) = \frac{1}{2\sqrt{\gamma \gamma_0}}f_{h_{fpl}}\Big(\sqrt{\frac{\gamma}{\gamma_0}}\Big)$. To derive the CDF, we integrate the PDF $ F_{h_{fpl}}(y) = \int_{0}^{y} f_{h_{fpl}}(y) dy $, to get the inner integral as $ \int_{0}^{y} y^{\alpha(1+A_1+A_2+s_1+s_2+s_3+s_4)} dy $. Solving the inner integral and applying the definition of Fox's H-function, the CDF is given in \eqref{eq:cdf_aekm_rpl}. Similar to the PDF, the CDF of the SNR can be derived by simple transformation of a random variable as $F_{\gamma}(\gamma) = \Big(\sqrt{\frac{\gamma}{\gamma_0}}\Big)$ to conclude the proof.

\section*{Appendix B}
The combined PDF of random path loss and antenna misalignment error $f_{h_{\rm lp}}(x)$ is given by \cite{papoulis_2002}
\begin{flalign} \label{eq:combined_hl_hp_eqn}
f_{h_{\rm lp}}(x) = \int _{x}^{a} \frac {1}{h_l} f_{h_{l}}(h_l)  f_{h_{p}}\left ({\frac {x}{h_l}}\right)  \mathrm {d}h_l.
\end{flalign}

Substituting the PDFs of the random path-loss and antenna misalignment error from  \eqref{eq:pdf_path_loss} and \eqref{eq:pdf_pointing_thz}, respectively, in \eqref{eq:combined_hl_hp_eqn}, to get the PDF as

\begin{flalign} \label{eq:combined_hl_hp_pdf_der}
f_{h_{lp}}(x) = - \frac{z^k \rho^2 x^{\rho-1}}{\Gamma(k)} \int _{x}^{a_l} h_{l}^{z-\rho -1} \bigg[ln\bigg(\frac{a_l}{h_{l}}\bigg)\bigg]^{k-1}    \ln\bigg(\frac{x}{h_l}\bigg)   \mathrm {d}h_l
\end{flalign}
substituting $\ln(\frac{a_l}{h_l}) = t$ in \eqref{eq:combined_hl_hp_pdf_der}, and applying the identity $\int_{0}^{u} x^{\nu-1} e^{-\mu x} dx = \mu^{-\nu}\big[\Gamma(\nu)-\Gamma(\nu,\mu u)\big]$ in \eqref{eq:combined_hl_hp_pdf_der}, we get the combined PDF of random path-loss and antenna misalignment error in \eqref{eq:pdf_hl_hp}. The PDF of SNR can be derived by simple transformation of a random variable \cite{papoulis_2002} as $f_{\gamma}(\gamma) = \frac{1}{2\sqrt{\gamma \gamma_0}}f_{h_{fpl}}\Big(\sqrt{\frac{\gamma}{\gamma_0}}\Big)$. To derive the CDF, we will integrate the PDF $F_{h_{lpl}}(y) = \int_{0}^{y} f_{h_{lpl}}(x) dx $ and use the series expansion of the upper incomplete Gamma function $\Gamma(a,z) = (a-1)! e^{-z}\sum_{j=0}^{a-1}\frac{z^j}{j!}$, to get
\begin{flalign} 
& F_{h_{lp}}(y) = - \frac{z^k \rho^2 a_l^{z-\rho}   (z-\rho)^{-k}}{\Gamma(k)} \nonumber \\ & \Bigg[ \frac{1}{z-\rho} \Big[ \int_{0}^{y}  x^{\rho-1} \Gamma(k+1)  dx \nonumber \\ &  -  k! \sum_{j=0}^{k} \frac{\Big((z-\rho)\Big)^j}{j!}  \int_{0}^{y}  x^{\rho-1}  \Big(\frac{a_l}{x}\Big)^{-(z-\rho)} \Big(\ln\big(\frac{a_l}{x}\big)\Big)^j   dx  \Big]  \nonumber \\ &  - \Big[ \int_{0}^{y}  x^{\rho-1} \Gamma(k) dx  - (k-1)! \nonumber \\ &  \sum_{j=0}^{k-1} \frac{\Big((z-\rho)\Big)^j}{j!} \int_{0}^{y}  x^{\rho-1}   \Big(\frac{a_l}{x}\Big)^{-(z-\rho)} \Big(\ln\big(\frac{a_l}{x}\big)\Big)^j  dx  \Big] \Bigg]
\end{flalign}
substituting $\ln(\frac{a_l}{x}) = t$ and applying the identity \cite[3.351,2]{Gradshteyn} $\int_{u}^{\infty} x^n e^{-\mu x} dx = \mu^{-n-1} \Gamma(n+1, \mu u)$, we get the combined CDF of random path-loss and antenna misalignment error in \eqref{eq:cdf_hl_hp}. The CDF of the SNR can be derived by transforming the random variable as $F_{\gamma}(\gamma) = \Big(\sqrt{\frac{\gamma}{\gamma_0}}\Big)$ to finish the proof.

\section*{Appendix C: Proof of  Lemma 1}
Since  $f(k)= 1/(k\cdot \exp({k\log r}))$ satisfies the integral inequality  $\sum_{k=1}^K f(k)\leq f(1)+\int_{1}^{K-1}f(x)dx$ in the interval of $1 \leq k \leq K-1$,  $D_{\rm{FTP}}$ in (\ref{eq:M_FTP}) can be upper bounded:
\begin{align}
	D_{\rm{FTP}}\leq (K-1)\Big(\frac{K}{K-1}+I_{K-1}+f(K)\Big),
	\label{main_first}
\end{align}
where $I_{K-1} = \int_{1}^{K-1}\frac{dx}{x\cdot\exp({x\log r})}$.
Substituting $z=x\log r$, we can express $I_{K-1}$:
\begin{align}
	I_{K-1} =  \Big(E_1(\log r)-E_1\big((K-1)\log r\big)\Big).
	\label{integral_first}
\end{align}
Using  series expansion of the exponential integral $E_1(z)=-\gamma-\log z-\sum_{m=1}^{\infty}{(-1)^mz^m}/{(m\cdot m!)} $ \cite[5.1.11]{Abramowitz1972book}, we can represent (\ref{integral_first}):
\begin{eqnarray}
	I_{K-1} = \log (K-1)+\sum_{m=1}^{\infty}\frac{\big((K-1)\log r\big)^{2m}}{2m\cdot 2m!}\hspace{-1mm}-\frac{(\log r)^{2m}}{2m \cdot 2m!} \hspace{-6.6mm}\nonumber \\+\sum_{m=1}^{\infty}\frac{\big(\log r\big)^{2m+1}}{2m+1\cdot 2m+1!}-\frac{((K-1)\log r)^{2m+1}}{2m+1 \cdot 2m+1!}
	\label{integral_third}
\end{eqnarray}
We use $\log(r)<-1/K$ and $\log(r)>-1/(K-1)$ in (\ref{integral_third})  appropriately to simplify:
\begin{align}
I_{K-1} \leq \log (K-1)+I_1 -I_2
\label{integral_sixth}
\end{align}
where $I_1= \sum_{m=1}^{\infty}{1}/{(m\cdot m!)} $ and  $I_2 =\sum_{m=0}^{\infty}{\big(1/K\big)^{m}}/{(m\cdot m!)} $.  Using the definition of the exponential integral $E_1(z)$, we  obtain  $I_1 = -(E_1(-1)+\gamma+i \pi )$. We also  derive an  upper bound $I_1\leq \sum_{m=1}^{\infty}{1}/{ m!} = \mathrm{e}-1$ with an error bound $\epsilon (I_1) <\mathrm{e}-J_0(2)$, where $J_0(\cdot)$ is the zeroth order Bessel function of the first kind. After a simple algebraic manipulation, we get a lower bound on $I_2>J_0\big(2/ \sqrt{K})\big)-1$. Using these  bounds in (\ref{integral_sixth}) with $\log (K-1) = \log K +\log (1-1/K)< \log K -1/K$:
\begin{align}
I_{K-1} \leq \log K -\frac{1}{K}+\mathrm{e} -J_0(2/\sqrt{K})
\label{IK_final}
\end{align}
Further, we represent $f(K) = 1/(K-1). \exp\big((K-1)\log (1+1/(K-1)\big)$ and  use $\log(1+1/(K-1)>1/(K-1)$ to bound  $f(K)$:
\begin{align}
f(K) < \frac{\mathrm{e}}{{K-1}}
\label{fk_final}
\end{align}
Thus, using  (\ref{IK_final}) and  (\ref{fk_final}) in (\ref{main_first}), we get the upper bound of Lemma 1.

 To get the the lower bound, we apply the upper bound $\log r < -\frac{1}{K}$ and subsequently the lower bound $\exp(\frac{k}{K}) > 1 + \frac{k}{K}$ in (\ref{eq:M_FTP}), we obtain $M_{\text{FTP}} > (K-1)(1+\mathcal{H}_K)$, where $\mathcal{H}_K = \sum_{k=1}^K \frac{1}{k}$ represents the $K$-th harmonic number.

Furthermore, employing a lower bound on $\mathcal{H}_K > \log (K+1/2) + \gamma$ \cite{Qi2014}, we can express $\log (K+1/2) = \log K + \log (1+1/2K)$. By utilizing the inequality $\log (1+x) > x/(1+x)$, we deduce $\log (1+1/2K) > 1/(1+2K)$, thereby obtaining another lower bound on $\mathcal{H}_K > \log K + 1/(1+2K) + \gamma$. Consequently, we establish the lower bound as stated in (\ref{lemma:ftp_bound}).

By employing the derived lower and upper bounds mentioned in (\ref{lemma:ftp_bound}), it is apparent that the scaling law remains valid as $K \rightarrow \infty$.

\bibliographystyle{IEEEtran}

\bibliography{jsac_bib_file}

\begin{thebibliography}{10}
\providecommand{\url}[1]{#1}
\csname url@samestyle\endcsname
\providecommand{\newblock}{\relax}
\providecommand{\bibinfo}[2]{#2}
\providecommand{\BIBentrySTDinterwordspacing}{\spaceskip=0pt\relax}
\providecommand{\BIBentryALTinterwordstretchfactor}{4}
\providecommand{\BIBentryALTinterwordspacing}{\spaceskip=\fontdimen2\font plus
\BIBentryALTinterwordstretchfactor\fontdimen3\font minus
  \fontdimen4\font\relax}
\providecommand{\BIBforeignlanguage}[2]{{%
\expandafter\ifx\csname l@#1\endcsname\relax
\typeout{** WARNING: IEEEtran.bst: No hyphenation pattern has been}%
\typeout{** loaded for the language `#1'. Using the pattern for}%
\typeout{** the default language instead.}%
\else
\language=\csname l@#1\endcsname
\fi
#2}}
\providecommand{\BIBdecl}{\relax}
\BIBdecl

\bibitem{Bhardwaj2024_wcnc}
P.~Bhardwaj \emph{et~al.}, ``A generalized statistical model for {THz} wireless
  channel with random atmospheric absorption,'' arXiv:2310.18616v1, Oct. 2023,
  submitted to 2024 IEEE Wireless Communications and Networking Conference
  ({IEEE WCNC} 2024).

\bibitem{Koenig_2013_nature}
S.~{Koenig} \emph{et~al.}, ``Wireless {sub-THz} communication system with high
  data rate,'' \emph{Nature Photon}, vol.~7, p. 977–981, 2013.

\bibitem{Dang_2020_nature}
S.~Dang \emph{et~al.}, ``What should {6G} be?'' \emph{Nature Electron}, no.~3,
  p. 20–29, 2020.

\bibitem{Ngo2017}
H.~Q. Ngo \emph{et~al.}, ``{Cell-Free} massive {MIMO} versus small cells,''
  \emph{IEEE Trans. Wireless Commun.}, vol.~16, no.~3, pp. 1834--1850, 2017.

\bibitem{Interdonato2019}
G.~Interdonato \emph{et~al.}, ``Ubiquitous cell-free massive {MIMO}
  communications,'' \emph{J Wireless Com Network}, vol. 197, 2019.

\bibitem{Zhang2019}
J.~Zhang \emph{et~al.}, ``Cell-free massive {MIMO}: A new next-generation
  paradigm,'' \emph{IEEE Access}, vol.~7, pp. 99\,878--99\,888, 2019.

\bibitem{Elhoushy2022}
S.~Elhoushy \emph{et~al.}, ``{Cell-Free} massive {MIMO}: {A} survey,''
  \emph{IEEE Commun. Surv. $\&$ Tut.}, vol.~24, no.~1, pp. 492--523, 2022.

\bibitem{Faisal2020_thz_cell_free}
A.~Faisal \emph{et~al.}, ``Ultramassive {MIMO} systems at {Terahertz} bands:
  Prospects and challenges,'' \emph{IEEE Veh. Technol. Mag.}, vol.~15, no.~4,
  pp. 33--42, 2020.

\bibitem{Sayyari2021_thz_cell_free}
R.~Sayyari \emph{et~al.}, ``Cell-free massive {MIMO} system with an adaptive
  switching algorithm between cooperative {NOMA}, non-cooperative {NOMA}, and
  {OMA} modes,'' \emph{IEEE Access}, vol.~9, pp. 149\,227--149\,239, 2021.

\bibitem{Abbasi2022_thz_cell_free}
O.~Abbasi and H.~Yanikomeroglu, ``A cell-free scheme for {UAV} base stations
  with {HAPS}-assisted backhauling in {Terahertz} band,'' in \emph{ICC 2022 -
  IEEE Int. Conf. Commun.}, 2022, pp. 249--254.

\bibitem{Mukherjee2022_thz_cell_free}
A.~Mukherjee, ``Jamming vulnerability of {Terahertz} wireless networks,'' in
  \emph{MILCOM 2022 - 2022 IEEE Military Commun. Conf. (MILCOM)}, 2022, pp.
  426--430.

\bibitem{Li2022_dual_hop_thz_fso}
S.~Li \emph{et~al.}, ``Mixed {THz/FSO} relaying systems: Statistical analysis
  and performance evaluation,'' \emph{IEEE Trans. Wireless Commun.}, vol.~21,
  no.~12, pp. 10\,996--11\,010, 2022.

\bibitem{Singya2022_hybrid_fso_thz_backhaul}
P.~K. Singya \emph{et~al.}, ``Hybrid {FSO/THz}-based backhaul network for
  {mmWave} terrestrial communication,'' \emph{IEEE Trans. Wireless Commun.},
  vol.~22, no.~7, pp. 4342--4359, 2023.

\bibitem{Pai2021_dual_hop_THz_backhaul}
V.~U. Pai \emph{et~al.}, ``Performance analysis of dual-hop {THz} wireless
  transmission for backhaul applications,'' in \emph{2021 IEEE Int. Conf. Adv.
  Netw. Telecommun. Systems (ANTS)}, 2021, pp. 438--443.

\bibitem{Li_2021_THz_AF}
S.~Li and L.~Yang, ``Performance analysis of dual-hop {THz} transmission
  systems over $\alpha$-$\mu $ fading channels with pointing errors,''
  \emph{IEEE Internet of Things J.}, pp. 1--1, 2021.

\bibitem{Bhardwaj2022_multihop}
P.~Bhardwaj and S.~M. Zafaruddin, ``On the performance of multihop {THz}
  wireless system over mixed channel fading with shadowing and antenna
  misalignment,'' \emph{IEEE Trans. Commun.}, pp. 1--1, 2022.

\bibitem{Ning2021_thz_multiuser}
B.~Ning \emph{et~al.}, ``Terahertz multi-user massive {MIMO} with intelligent
  reflecting surface: Beam training and hybrid beamforming,'' \emph{IEEE Trans.
  Veh. Technol.}, vol.~70, no.~2, pp. 1376--1393, 2021.

\bibitem{Chen2023_thz_multiuser}
R.~Chen \emph{et~al.}, ``Multi-user orbital angular momentum based {Terahertz}
  communications,'' \emph{IEEE Trans. Wireless Commun.}, pp. 1--1, 2023.

\bibitem{Wang2023_thz_multiuser}
Y.~Wang \emph{et~al.}, ``Sensing-aided hybrid precoding for efficient
  {Terahertz} wideband communications in multi-user high-data-rate {IoT},''
  \emph{TechRxiv}, June 2023.

\bibitem{Chen2020_thz_multiuser}
C.~Wenjie \emph{et~al.}, ``Channel estimation for intelligent reflecting
  surface aided multi-user {MISO} {Terahertz} system,'' \emph{Terahertz Sci.
  $\&$ Technol.}, vol.~13, no.~2, pp. 51--60, 2020.

\bibitem{Lee2022_thz_random_access}
S.~Lee \emph{et~al.}, ``A new preamble signal design for random access in
  sub-{Terahertz} {6G} cellular systems,'' in \emph{2022 IEEE Int. Conf.
  Commun. Workshops (ICC Workshops)}, 2022, pp. 1147--1152.

\bibitem{Emil2020}
E.~Björnson and L.~Sanguinetti, ``Scalable {Cell-Free} massive {MIMO}
  systems,'' \emph{IEEE Trans. Commun.}, vol.~68, no.~7, pp. 4247--4261, 2020.

\bibitem{Liu2020}
P.~Liu \emph{et~al.}, ``Spectral efficiency analysis of {Cell-Free} massive
  {MIMO} systems with zero-forcing detector,'' \emph{IEEE Trans. Wireless
  Commun.}, vol.~19, no.~2, pp. 795--807, 2020.

\bibitem{Emil2020twc}
E.~Björnson and L.~Sanguinetti, ``Making {Cell-Free} massive {MIMO}
  competitive with {MMSE} processing and centralized implementation,''
  \emph{IEEE Trans. Wireless Commun.}, vol.~19, no.~1, pp. 77--90, 2020.

\bibitem{Zhang2021}
J.~Zhang \emph{et~al.}, ``Local partial zero-forcing combining for {Cell-Free}
  massive {MIMO} systems,'' \emph{IEEE Trans. Commun.}, vol.~69, no.~12, pp.
  8459--8473, 2021.

\bibitem{Ibrahim2022}
M.~Ibrahim \emph{et~al.}, ``Uplink performance of {MmWave}-fronthaul
  {Cell-Free} massive {MIMO} systems,'' \emph{IEEE Trans. Veh. Technol.},
  vol.~71, no.~2, pp. 1536--1548, 2022.

\bibitem{Rivero2005}
M.~Rivero-Angeles \emph{et~al.}, ``Random-access control mechanisms using
  adaptive traffic load in {ALOHA} and {CSMA} strategies for {EDGE},''
  \emph{IEEE Trans. Veh. Technol.}, vol.~54, no.~3, pp. 1160--1186, 2005.

\bibitem{Baccelli2013}
F.~Baccelli and C.~Singh, ``Adaptive spatial {ALOHA}, fairness and stochastic
  geometry,'' in \emph{2013 11th Int. Symp. Workshops Modeling Optim. in
  Mobile, Ad Hoc and Wireless Netw. (WiOpt)}, 2013, pp. 7--14.

\bibitem{Wang2016}
L.~Wang \emph{et~al.}, ``Adaptive-opportunistic {Aloha}: A media access control
  protocol for unmanned aerial vehicle–wireless sensor network systems,''
  \emph{Int. J. Distrib. Sensor Netw.}, vol.~12, no.~8, p. 1550147716662785,
  2016.

\bibitem{Vukobratovic2020}
D.~Vukobratovic and F.~J. Escribano, ``Adaptive multi-receiver coded slotted
  {ALOHA} for indoor optical wireless communications,'' \emph{IEEE Commun.
  Lett.}, vol.~24, no.~6, pp. 1308--1312, 2020.

\bibitem{Zhang2020}
M.~Zhang \emph{et~al.}, ``Adaptive policy tree algorithm to approach
  collision-free transmissions in slotted {ALOHA},'' in \emph{2020 IEEE 17th
  Int. Conf. Mobile Ad Hoc Sensor Syst. (MASS)}, 2020, pp. 138--146.

\bibitem{Ebrahimi2023}
M.~M. Ebrahimi \emph{et~al.}, ``Adaptive–persistent nonorthogonal random
  access scheme for {URLL} massive {IoT} networks,'' \emph{IEEE Syst. J.},
  vol.~17, no.~1, pp. 1660--1671, 2023.

\bibitem{rom90book}
R.~Rom and M.~Sidi, \emph{Multiple Access Protocols: Performance and
  Analysis}.\hskip 1em plus 0.5em minus 0.4em\relax Springer-Verlag, 1990.

\bibitem{Schenk_hw}
T.~Schenk, \emph{{RF} Imperfections in High-Rate Wireless Systems: Impact and
  Digital Compensation}.\hskip 1em plus 0.5em minus 0.4em\relax Springer, 2008.

\bibitem{Boulogeorgos_Analytical}
A.~A. {Boulogeorgos} and A.~{Alexiou}, ``Analytical performance assessment of
  {THz} wireless systems,'' \emph{IEEE Access}, vol.~7, pp. 11\,436--11\,453,
  2019.

\bibitem{Pranay_2021_TVT}
P.~Bhardwaj and S.~M. Zafaruddin, ``Performance of dual-hop relaying for
  {THz-RF} wireless link over asymmetrical $\alpha$-$\mu$ fading,'' \emph{IEEE
  Trans. Veh. Technol.}, vol.~70, no.~10, pp. 10\,031--10\,047, 2021.

\bibitem{Bhardwaj2022_systems_journal}
------, ``Performance of hybrid {THz} and multiantenna {RF} system with
  diversity combining,'' \emph{IEEE Syst. J.}, pp. 1--12, 2022.

\bibitem{Joshi2022_ftr}
A.~A. Joshi \emph{et~al.}, ``Terahertz wireless transmissions with maximal
  ratio combining over {Fluctuating Two-Ray} fading,'' in \emph{2022 IEEE
  Wireless Commun. Netw. Conf. (WCNC)}, 2022, pp. 1575--1580.

\bibitem{Bhardwaj2023_outdoor_thz}
P.~Bhardwaj and S.~M. Zafaruddin, ``Performance analysis of outdoor {THz}
  wireless transmission over mixed gaussian fading with pointing errors,'' in
  \emph{Int. Conf. Next Gener. Syst. Netw.}, 2023, pp. 1--1.

\bibitem{Bhardwaj2023_iot}
P.~Bhardwaj \emph{et~al.}, ``Performance of integrated {IoT} network with
  hybrid {mmWave/FSO/THz} backhaul link,'' \emph{IEEE Internet Things J.}, pp.
  1--1, 2023.

\bibitem{Bhardwaj2023_THI}
P.~Bhardwaj and S.~M. Zafaruddin, ``Exact performance analysis of {THz} link
  under transceiver hardware impairments,'' \emph{IEEE Commun. Lett.}, pp.
  1--1, 2023.

\bibitem{Papasotiriou2021_scientific_report}
E.~Papasotiriou \emph{et~al.}, ``An experimentally validated fading model for
  {THz} wireless systems,'' \emph{Sci. report}, vol.~11, 2021.

\bibitem{Papasotiriou2023_scintific_report_outdoor}
------, ``Outdoor {THz} fading modeling by means of gaussian and gamma mixture
  distributions,'' \emph{Scientific Rep.}, vol.~13, no. 6385, 2023.

\bibitem{Yacoub_2016_alpha_eta_kappa_mu}
M.~D. Yacoub, ``The $\alpha $ - $\eta $ - $\kappa $ - $\mu $ fading model,''
  \emph{IEEE Trans. Ant. Prop.}, vol.~64, no.~8, pp. 3597--3610, 2016.

\bibitem{Marins2019_alpha_eta_kapp_mu}
T.~R.~R. Marins \emph{et~al.}, ``Fading evaluation in the {mm-Wave} band,''
  \emph{IEEE Trans. Commun.}, vol.~67, no.~12, pp. 8725--8738, 2019.

\bibitem{Bhardwaj2023_alpha_eta_kappa_mu_globecom}
P.~Bhardwaj \emph{et~al.}, ``An exact statistical representation of
  $\alpha$-$\eta$-$\kappa$-$\mu$ fading model for {THz} wireless
  communication,'' Accepted in 2023 IEEE Globecom Workshops (GC 2023 Workshop -
  HCWC), Kuala Lumpur, Malaysia, p.1-6.

\bibitem{Dabiri2022_zaf}
M.~T. Dabiri and M.~Hasna, ``Pointing error modeling of {mmWave} to {THz}
  high-directional antenna arrays,'' \emph{IEEE Wireless Communications
  Letters}, vol.~11, no.~11, pp. 2435--2439, 2022.

\bibitem{Badarneh2023_THz_Pointing}
O.~S. Badarneh \emph{et~al.}, ``Channel modeling and performance analysis of
  directional {THz} links under pointing errors and $\alpha$-$\mu$
  distribution,'' \emph{IEEE Commun. Lett.}, pp. 1--1, 2023.

\bibitem{Kim2015}
S.~{Kim} and A.~G. {Zajić}, ``Statistical characterization of {300-GHz}
  propagation on a desktop,'' \emph{IEEE Trans. Veh. Technol.}, vol.~64, no.~8,
  pp. 3330--3338, 2015.

\bibitem{Kokkoniemi_2018}
J.~{Kokkoniemi} \emph{et~al.}, ``Simplified molecular absorption loss model for
  275–400 {Gigahertz} frequency band,'' in \emph{12th Eur. Conf. Antenna
  Propag. (EuCAP 2018)}, 2018, pp. 1--5.

\bibitem{Wu2020}
Y.~{Wu} \emph{et~al.}, ``Interference and coverage analysis for {Terahertz}
  networks with indoor blockage effects and line-of-sight access point
  association,'' \emph{IEEE Trans. Wireless Commun.}, vol.~20, no.~3, pp.
  1472--1486, 2021.

\bibitem{fog}
M.~A. Esmail \emph{et~al.}, ``On the performance of optical wireless links over
  random foggy channels,'' \emph{IEEE Access}, vol.~5, pp. 2894--2903, 2017.

\bibitem{Du2020_RIS_THz_HW_Impaiment}
H.~Du \emph{et~al.}, ``Performance and optimization of reconfigurable
  intelligent surface aided {THz} communications,'' \emph{IEEE Trans. Commun.},
  vol.~70, no.~5, pp. 3575--3593, 2022.

\bibitem{Abramson1970}
N.~Abramson, ``The {ALOHA system}: Another alternative for computer
  communications,'' in \emph{Proc. Fall Joint Computer Conf.}, 1970.

\bibitem{Abramowitz1972book}
M.~Abramowitz and I.~A. Stegun, \emph{Handbook of Mathematical Functions with
  Formulas, Graphs, and Mathematical Tables}, 10th~ed.\hskip 1em plus 0.5em
  minus 0.4em\relax Academic, 1972.

\bibitem{AboRahama_2018}
A.~Rahama \emph{et~al.}, ``On the sum of independent {Fox's H} - function
  variates with applications,'' \emph{IEEE Trans. Veh. Technol.}, vol.~67,
  no.~8, pp. 6752--6760, 2018.

\bibitem{Kilbas_2004}
A.~A. {Kilbas}, \emph{H-Transforms: Theory and Applications}.\hskip 1em plus
  0.5em minus 0.4em\relax CRC Press, 2004, vol. First edition.

\bibitem{biswas2015}
K.~Biswas, V.~Muthukkumarasamy, X.~W. Wu, and K.~Singh, ``An analytical model
  for lifetime estimation of wireless sensor networks,'' \emph{IEEE
  Communications Letters}, vol.~19, no.~9, pp. 1584--1587, Sept 2015.

\bibitem{papoulis_2002}
{A. Papoulis} and {S. Pillai}, \emph{Probability, Random Variables, and
  Stochastic Processes}.\hskip 1em plus 0.5em minus 0.4em\relax McGraw Hill,
  Boston, Fourth Edition, 2002.

\bibitem{Gradshteyn}
I.~S. {Gradshteyn} and I.~M. {Ryzhik }, \emph{Table of Integrals, Series, and
  Products}.\hskip 1em plus 0.5em minus 0.4em\relax Academic press, San Diego,
  CA, 6th edition, 2000.

\bibitem{Qi2014}
B.-N. Guo and F.~Qi, ``Sharp inequalities for the psi function and harmonic
  numbers,'' \emph{Analysis-International mathematical journal of analysis and
  its applications}, vol.~34, no.~2, pp. 201--208, June 2014.

\end{thebibliography}

\end{document}